\documentclass[conference,compsoc]{IEEEtran}
\IEEEoverridecommandlockouts % for using \thanks
\usepackage{courier}
\usepackage{SIunits}
\usepackage[bookmarks=false,colorlinks=true,breaklinks,draft=false,linkcolor=blue,citecolor=blue,filecolor=blue,urlcolor=blue]{hyperref}
\usepackage[cmex10]{amsmath}
\usepackage{mathtools}
\DeclarePairedDelimiter\ceil{\lceil}{\rceil}

\usepackage{graphicx}
\usepackage{url}
\usepackage[caption=false,font=footnotesize]{subfig}
\usepackage{listings,verbatim}
\usepackage{multirow}
\usepackage{array}
\usepackage{xcolor}
\usepackage{tikz}
\usetikzlibrary{shapes,arrows,backgrounds}
\usetikzlibrary{shapes.multipart}
\usetikzlibrary{positioning}
\DeclareMathAlphabet{\mathcal}{OMS}{cmsy}{m}{n}
\newcolumntype{P}[1]{>{\centering\arraybackslash}p{#1}}
\newcommand{\F}{Fig.}

\newcommand{\ignore}[1]{}
\newcommand{\shortlong}[2]{#1}
\renewcommand{\shortlong}[2]{#2}

\newcommand{\longversion}[1]{\shortlong{}{#1}}

\newcommand{\babel}[0]{\textsc{Babel}}
\newcommand{\mo}{translingual obfuscation}
\newcommand{\Mo}{Translingual obfuscation}
\newcommand{\MO}{Translingual Obfuscation}
\newcommand{\MOlb}{Translingual\\Obfuscation}

\newcommand{\vbo}{virtualization-based obfuscation}

\begin{document}

\title{\MO\longversion{\thanks{This is an extended version of a paper to appear in
 \textit{Proceedings of the 1st IEEE European Symposium on Security and Privacy (Euro S\&P 2016)} \cite{mo}.}}}

\author{
\IEEEauthorblockN{Pei Wang, Shuai Wang, Jiang Ming, Yufei Jiang, and Dinghao Wu}
\IEEEauthorblockA{College of Information Sciences and Technology\\
The Pennsylvania State University\\
\{pxw172, szw175, jum310, yzj107, dwu\}@ist.psu.edu}
}
\maketitle

\begin{abstract}
  Program obfuscation is an important software protection technique
  that prevents attackers from revealing the programming logic and design of the software.
  We introduce \emph{\mo}, a new software obfuscation scheme which
  makes programs obscure by ``misusing'' the unique features of certain programming languages. 
  \Mo\ translates part of a program from its original language
  to another language which has a different programming paradigm and execution model,
  thus increasing program complexity and impeding reverse engineering.
  In this paper, we investigate the feasibility and effectiveness of \mo\ with Prolog,
  a logic programming language.
  We implement \mo\ in a tool called \babel, which can selectively translate C functions into
  Prolog predicates.
  By leveraging two important features of the Prolog language,
  i.e., unification and backtracking, \babel\ obfuscates both the data layout and control flow of 
  C programs, making them much more difficult to reverse engineer.
  Our experiments show that \babel\ provides effective
  and stealthy software obfuscation, while the cost is only modest compared to one of
  the most popular commercial obfuscators on the market.
  With \babel, we verified the feasibility of \mo, which we consider to be
  a promising new direction for software obfuscation. 
\end{abstract}

\section{Introduction}\label{sec:intro}
Obfuscation is an important technique for software protection,
especially for preventing reverse engineering from infringing software intellectual property. 
Generally speaking, obfuscation is a semantics-preserving
program transformation that aims to make a program
more difficult to understand and reverse engineer.
The idea of using obfuscating transformations to prevent
reverse engineering can be traced back to 
 Collberg et al.~\cite{Collberg1997,Collberg1998, nagra2009surreptitious}. Since then many obfuscation
methods have been
proposed~\cite{Linn2003,Moser2007,Popov2007,Sharif2008,Chen:2009:CFO:1669112.1669162,Wu:2010:MNA:1866307.1866368}.%
\longversion{
Malware authors also heavily rely on obfuscation to compress or
encrypt executable binaries so that their products can avoid malicious content
detection~\cite{Szor2005,Sikorski2012}.
}

Currently the state-of-the-art obfuscation technique is to incorporate with
\textit{process-level virtualization}.
For example, obfuscators such as VMProtect~\cite{VMProtect} and Code
Virtualizer~\cite{code-virtualizer} replace the original binary code
with new bytecode, and a custom interpreter is attached to
interpret and execute the bytecode. The result is that the original
binary code does not exist anymore, leaving only the bytecode and
interpreter, making it difficult to directly reverse engineer~\cite{proceeding_Justin_RAID08}.
However, recent work has shown that the decode-and-dispatch execution pattern 
of \vbo\ can be a severe vulnerability leading to effective deobfuscation~\cite{Coogan2011,Sharif2009},
implying that we are in need of obfuscation techniques based on new schemes.

We propose a novel and practical obfuscation method called \textit{\mo}, 
which possesses strong security strength and good stealth, with only modest cost.
The key idea is that instead of inventing brand new obfuscation techniques,
we can exploit some existing programming languages for their unique design and 
implementation features to achieve obfuscation effects.
In general, programming language features are rarely proposed or developed for obfuscation purposes;
however, some of them indeed make reverse engineering much more challenging at the binary level and
thus can be ``misused'' for software protection.
In particular, some programming languages are designed with unique paradigms and have very complicated execution models.
To make use of these language features, we can translate a program
written in a certain language to another language which is more ``confusing'',
in the sense that it consists of features leading to obfuscation effects.

In this paper, we obfuscate C programs by translating them into Prolog,
presenting a feasible example of the \mo{} scheme.
C is a traditional imperative programming language while Prolog is
a typical logic programming language.
The Prolog language has some prominent features that provide strong obfuscation effects.
Programs written in Prolog are executed in a \textit{search-and-backtrack} computation model
which is dramatically different from the execution model of C and much more complicated.
Therefore, translating C code to Prolog leads to obfuscated data layouts and control flows.
Especially, \emph{the complexity of Prolog's execution model manifests mostly in the binary
form of the programs,} making Prolog very suitable for software protection.

Translating one language to another is usually very difficult,
especially when the target and source languages have different programming paradigms.
However, we made an important observation that for obfuscation purposes, 
language translation could be conducted in a special manner.
Instead of developing a ``clean'' translation from C to Prolog,
we propose an ``obfuscating'' translation scheme which retains part of the C memory model,
in some sense making two execution models mixed together.
We believe this improves the obfuscating effect in a way that no obfuscation methods 
have achieved before, to the best of our knowledge.
Consequently in \mo, the obfuscation does not only come from the obfuscating features of 
the target language, but also from the translation itself.
With this new translation scheme we manage to kill two birds with one stone, i.e.,
solving the technical problems in implementing \mo\ and strengthening the obfuscation
simultaneously. 

There may be of a concern that obfuscation techniques without solid theoretical 
foundations will not withstand reverse engineering attacks in the long run. However,
research on fundamental obfuscation theories, despite promising process made
recently~\cite{\shortlong{lopez-alt_--fly_2012,garg_candidate_2013,
sahai_how_2014}{lopez-alt_--fly_2012,garg_candidate_2013,
sahai_how_2014,boyle_extractability_2014,barak_protecting_2014}},
is still not mature enough to spawn practical protection techniques.
There is a widely accepted consensus that no software protection scheme is resilient to
skilled attackers if they inspect the software with intensive effort~\cite{collberg_watermarking_2002}.
A recently proved theorem~\cite{barak_impossibility_2012} partially supporting this claim states that, a ``universally effective'' obfuscator
does not exist, i.e., for any obfuscation algorithm, there always exists an program that it cannot
effectively obfuscate.
Given the situation, it seems that developing an obfuscation scheme resilient to all
reverse engineering threats (known or unknown) is too ambitious at this point. Hence, 
making reverse engineering more difficult (but not impossible) could be a more realistic 
goal to pursue.

We have implemented \mo\ in a tool called \babel. \babel\ can
selectively transform a C function into semantically equivalent Prolog code and
compile code of both languages together into the executable form.
Our experiment results show that \mo\ is obscure and
stealthy. 
The execution overhead of \babel\ is modest compared to a commercial obfuscator.
We also show that
\mo\ is resilient to one of the most popular reverse engineering techniques.

In summary, we make the following contributions in this research:
\begin{itemize}
\item We propose a new obfuscation method, namely \mo.
  \Mo\ is novel because it exploits exotic language features instead of ad-hoc 
  program transformations to 
  protect programs against reverse engineering.
  Our new method has a number of advantages over existing obfuscation techniques,
  which will be discussed in depth later.
\item We implement \mo\ in a tool called \babel\
  which translates C to Prolog at the scale of subroutines, i.e., from C functions
  to Prolog predicates, to obfuscate the original programs.
  Language translation is always a challenging problem, especially when the target language has
  a heterogeneous execution model.
\item We evaluate \babel\ with respect to all four evaluation criteria proposed by Collberg 
  et al.~\cite{Collberg1998}: potency, resilience, cost, and stealth, on a set of
  real-world C programs with quite a bit of complexity and diversity.
  Our experiments demonstrate that \babel\ provides strong protection against reverse engineering
  with only modest cost. %making \mo\ a feasible and promising new direction of software obfuscation.
\end{itemize}

The remainder of this paper is organized as follows. 
\longversion{
\S\ref{sec:tm} defines our threat model.
}
\S\ref{sec:mo} provides a high-level view on the insights and features of our \mo\ technique.
\S\ref{sec:background} explains in detail why the Prolog programming language 
can be misused for obfuscation.
We summarize the technical challenges in implementing \mo\ in \S\ref{sec:cpob}.
\S\ref{sec:design} and \S\ref{sec:impl}
present our C-to-Prolog translation method and 
the implementation details of \babel, respectively.
We evaluate \babel's performance in \S\ref{sec:eval}.
\S\ref{sec:discussion} has a discussion on some important topics about \mo,
followed by the summary of related work in \S\ref{sec:related}.
\S\ref{sec:conclusion} concludes the paper.

\longversion{
\section{Threat Model}\label{sec:tm}
For attackers who try to reverse engineer a program protected by obfuscation,
we assume that they have full access to the binary form of the program.
They can examine the static form of the binaries with whatever method available to them.
They can also execute the victim binaries in a monitored environment with arbitrary input,
thus can read any data that has lived in the memory. 

Do note that although we assume attackers have unlimited access to program binaries,
they should not posses any knowledge about the source code in our threat model.
Assuming attackers can only examine the obfuscated program at the binary level is important,
because that would mean any implementation detail of the language used in \mo\ contributes to
the effectiveness of obfuscation. As for the particular case of employing Prolog in \mo, 
since Prolog is a declarative programming language,
there is a much deeper semantic gap between its source code and binaries,
which is highlighted as one of the major sources of \mo's protection effects. 

Finally, we explicitly clarify that in this work,
attackers are assumed to try to reveal the logical structure
of the binaries so that they can reproduce the algorithms by themselves. 
In practice there are different levels of reverse engineering objectives. Sometimes
understanding what a program achieves is sufficient for attackers to fulfill their goals,
but in our case attackers need a more thorough understanding on the semantics
of the victim binaries.

Our threat model may seem too coarsely defined.
However, we believe it is quite realistic,
since reverse engineering could be a very ad-hoc process in practice.
Actually, lack of specifications makes it difficult for us to 
design and evaluate a new obfuscation technique in a fully comprehensive manner,
because we cannot make further assumptions on the methods or tools that
attackers may make use of.
Therefore,
we hope that readers of this paper could pay more attention to the general idea and picture we want to present.

}
\section{\MO}\label{sec:mo}
\subsection{Overview}

The basic idea of \mo\ is that some programming languages are more
difficult to reverse engineer than others. Intuitively, C is relatively
easy to reverse engineer because binary code compiled from C programs
shares the same imperative execution model with the source code.
For some programming languages like Prolog, however, there is a much deeper
gap between the source code and the
resulting binaries, since these languages have fundamentally different
abstractions from the imperative execution model of the underlying hardware.
Starting from this insight, we analyze and evaluate the
features of a foreign programming language from the perspective of software protection.
We also develop the translation technique that transforms the original language
to the obfuscating language. Only with these efforts devoted, \mo\ can be
a practical software protection scheme. 

\begin{figure}[t]
  \centering
  \begin{tikzpicture}[scale=0.95,every node/.style={transform shape}]
    \scriptsize
    \node[rectangle, fill=green!30, minimum width=2.5cm,minimum height=1.7cm] (sourceAO1) {};
    \coordinate[above=.25cm of sourceAO1.north east] (sAOa1);
    \coordinate[below=.25cm of sourceAO1.south east] (sAOb1);
    \draw[dashed] (sAOa1) -- (sAOb1);
    \node[rectangle, fill=green!30, minimum width=1.95cm, minimum height=1.7cm, right=0 of sourceAO1] (bin-ob1) {};
    \node[rectangle, fill=red!30, minimum width=4.45cm, minimum height=1.7cm, right=0 of bin-ob1] (bin-de1) {};

    \coordinate[above=.5cm of bin-ob1.north east] (ba1);
    \coordinate[below=.5cm of bin-ob1.south east] (bb1);
    \draw (ba1) -- (bb1);

    \node[tape, draw=black, fill=white,text width=2cm, text centered, inner sep = .2ex, below=-.1cm of sourceAO1.center] (sourceA1) {Source code\\in language $\mathcal{A}$};
    \node[above=2ex of sourceAO1.center, text width=2cm, text centered,text=black!100] {\sf Source-Code\\Obfuscation};

    \node[tape, draw = black, fill = white, text width = 2cm, text centered, inner sep = .2ex, minimum height=.8cm, below=-.1cm of bin-ob1.east,] (bin1) {Binary};

    \draw[->] (sourceA1) -- (bin1);

    \node[above=2ex of bin-ob1.center, text width=2cm, text centered,text=black!100] {\sf Binary\\Obfuscation};
    \node[above=3ex of bin-de1.center, text width=2cm, text centered,text=black!100] {\sf Deobfuscation};    

    \node[above = .1cm of sAOa1, xshift=-.5cm] (label1) {\textsf{\textbf{Developer Side}}};
    \node[right = 3cm of label1] {\textsf{\textbf{Attacker Side}}};

    \node[rectangle, fill=green!30, minimum width=2.5cm,minimum height=1.7cm, below=1cm of sourceAO1] (sourceAO) {};
    \coordinate[above=.25cm of sourceAO.north east] (sAOa);
    \coordinate[below=.25cm of sourceAO.south east] (sAOb);
    \draw[dashed] (sAOa) -- (sAOb);

    \node[rectangle, fill=green!90, right=0 of sourceAO, minimum width=3cm,minimum height=1.7cm] (sourceBO) {};

    \node[rectangle, fill=green!30, minimum width=1.7cm, minimum height=1.7cm, right=0 of sourceBO] (bin-ob) {};
    \node[rectangle, fill=red!30, minimum width=1.7cm, minimum height=1.7cm, right=0 of bin-ob] (bin-de) {};

    \coordinate[above=.5cm of bin-ob.north east] (ba);
    \coordinate[below=.5cm of bin-ob.south east] (bb);
    \draw (ba) -- (bb);

    \node[tape, draw=black, fill=white,text width=2cm, text centered, inner sep = .2ex, below=-.1cm of sourceAO.center] (sourceA) {Source code\\in language $\mathcal{A}$};

    \node[tape, draw=black, fill=white,text width=2.5cm, text centered, inner sep = .2ex, below=-.1cm of sourceBO.center] (sourceB) {Source code\\in language $\mathcal{B}$ (and $\mathcal{A}$)};

    \node[above=2ex of sourceAO.center, text width=2cm, text centered,text=black!100] {\sf Source-Code\\Obfuscation};

    \node[above=2ex of sourceBO.center, text width=2cm, text centered] {\textsf{\textbf{\MOlb}}};

    \node[tape, draw=black, fill=white, text width=2cm, minimum height=.8cm,text centered, inner sep=.2ex, below=-.1cm of bin-ob.east] (bin) {Binary};

    \draw[->] (sourceA) -- (sourceB);
    \draw[->] (sourceB) -- (bin);

    \node[above=2ex of bin-ob.center, text width=2cm, text centered,text=black!100] {\sf Binary\\Obfuscation};
    \node[above=3ex of bin-de.center, text width=2cm, text centered,text=black!100] {\sf Deobfuscation};    

%    \node[above = .1cm of sAOa] (label1) {\textsf{\textbf{Developer Side}}};
%    \node[right = 3.8cm of label1] {\textsf{\textbf{Attacker Side}}};

    \draw[->,line width=1pt] (bb1) -- (ba);

    \coordinate[above=.25cm of sourceBO.north east] (sAOa);
    \coordinate[below=.25cm of sourceBO.south east] (sAOb);
    \draw[dashed] (sAOa) -- (sAOb);
  \end{tikzpicture}
  \caption{\Mo\ is a new protection layer complementary to existing obfuscation methods, pushing the frontier forward in the battle with reverse engineering.}
  \label{fig:race}
\end{figure}
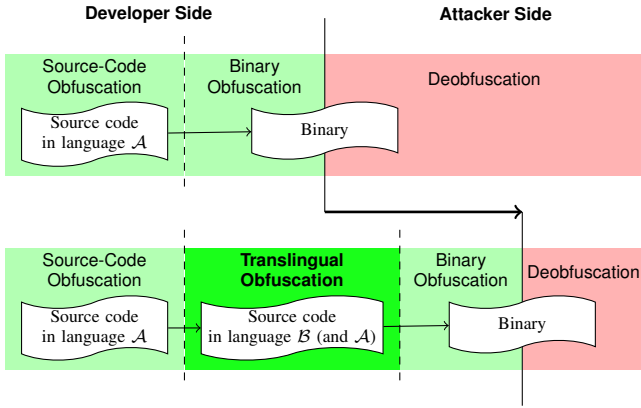

We view \mo\ as a new layer of software protection in the obfuscation-deobfuscation arms race,
as shown in \F~\ref{fig:race}.
%Normally we require that the language used by \mo\ can be directly compiled
%to native code.
Different from previous obfuscation methods which either work at the 
binary level or perform same-language source-to-source transformations, 
\mo\ translates one language to another.
Therefore, \mo\ can be applied after source-code obfuscation and before 
binary obfuscation without affecting the applicability of existing
obfuscation methods.

\subsection{Comparing with Virtualization-Based Obfuscation}
\label{sec:discussion:vo}
\begin{figure}[t]
  \centering
  \begin{tikzpicture}[trim right = 0cm, trim left=0cm]
    \node[circle,draw,minimum width=14em] (a) at (-1.4,0) {};
    \node[circle,draw,minimum width=14em] (b) at (1.4,0) {};
    \node[text width=10em,text centered] at (-1.4,3) {\footnotesize \textsf{\textbf{\MOlb\\}}};
    \node[text width=20em,text centered] at (1.4,3) {\footnotesize \textsf{\textbf{Virtualization-Based\\Obfuscation\\}}};
    \node[text width=9em, text centered] at (0,0) {\scriptsize \textsf{{Exotic Language\\Features\\+\\Virtualization\\}}};
    \node[circle,draw,minimum width=6.5em] (c) at (-2.15,-1) {};
    \node[circle,draw,minimum width=6.5em] (d) at (2.15,-1) {};
    \node[text width=6em, text centered] (e) at (-2.15,-1) {\scriptsize\textsf{\babel}\\\textsf{{(GNU Prolog,\\native code and \\no interpreter)\\}}};
    \node[text width=6em, text centered] (f) at (2.15,-1) {\scriptsize \textsf{{VMProtect, Code Virtualizer,\\etc.\\}}};
    \node[text width=6em, text centered] (g) at (-2.1,1) {\scriptsize\textsf{Exotic Language Features\\(heterogeneous\\programming paradigms)\\}};
    \node[text width=6em, text centered] (g) at (2.1,1) {\scriptsize\textsf{Virtualization\\(byte code interpretation)\\}};
  \end{tikzpicture}%
  \caption{Comparing \mo\ and \vbo.}
  \label{fig:relationship}
\end{figure}

The {\vbo} is currently the state of the art in binary obfuscation.
Some features of {\mo} resemble the idea of {\vbo}, but we want to emphasize a significant difference here.
% We have a brief comparison between \mo\ and \vbo\ in the introduction. 
%Another distinct new features of {\mo} is the mix of heterogeneous execution models and programming paradigms.
Currently, most implementations of {\vbo}
tend to encode original native machine code with a RISC-like virtual instruction set and interpret 
the encoded binary in a decode-dispatch pattern~\cite{smith05, Rolles:2009:UVO:1855876.1855877, Ghosh:2012:RAA:2151024.2151051},
which has been identified as a notable weakness of security and can be exploited
by various attacks~\cite{Sharif2009,Coogan2011,yadegari2014generic}.
%The obfuscation strength comes from the secrecy of encoding and interpretation details. 
%Since the target language is also imperative.
% \Vbo\ does not introduce a different execution model when protecting C programs.
% \footnote{Even if the original binaries are compiled from non-imperative languages such as Haskell and Ocaml,
% \vbo\ does not combine different programming paradigms like \mo\ does. 
% In this case, two different paradigms are not parallel mixed together but nested. This is a subtle but significant difference.}
\Mo, however, gets most of its security strength by
\textit{intentionally} relying on obfuscation-contributing language 
features that comes from a heterogeneous programming model.
Essentially, \mo\ does not have to re-encode the original binary as long
as the foreign language employed supports compilation into native code.
\F~\ref{fig:relationship} shows the relationship and key differences between the two methods.
Our \mo\ implementation \babel\ and {\vbo} do not overlap.

%Most importantly, as we showed in \F~\ref{fig:race}, \mo\ is not a direct competitor against \vbo, which is a binary obfuscation technique. They can be mixed together and cooperate for stronger obfuscation. 

\subsection{Benefits}
\label{sec:benefits}
\Mo\ can provide benefits
that cannot be delivered by any single obfuscation method developed before,
to the best of our knowledge:
\begin{itemize}
%\item The security strength of \mo\ could be as strong as \vbo\ at least, 
%which is currently the state of the art.
%Since many programming languages implement their runtime systems as virtual machines,
%\mo\ are comparable to \vbo\ in the aspect of obfuscation effectiveness. 
\item
\Mo{} provides strong obfuscation strength and more obfuscation variety by introducing a different programming paradigm.
If there exists \textit{a universally effective and automated method}
%% \footnote{As mentioned earlier,
%% for every obfuscation method, there exists a program such that a reverse engineering method,
%% maybe with manual effort, can effectively deobfuscate that particular obfuscated program.}
to nullify the obfuscation effects, namely the additional
program complexity, introduced by a programming language's execution model,
that would mean it is possible to significantly simplify the design and
implementation of that language,
which is very unlikely for mature languages.

\item \Mo\ can be very stealthy, because programming with multiple languages
is a completely legit practice.
Compared with \vbo\ which
encodes native code into bytecode that has an exotic encoding format, \mo\
introduces neither abnormal byte entropy nor deviant instruction distributions.
%weak against detection based on byte statistics~\cite{entropy07}.
% without additional cover-up effort~\cite{Wu:2010:MNA:1866307.1866368},
%\mo\ does not introduce byte statistical anomaly.
%Actually, programming with multiple languages is a completely legit practice. 

%\item The translation technique employed in \mo\ leads to a mixture of runtime features from
%both languages, which contributes to the obfuscation in general. 
%The rationale behind our ``obfuscating'' translation scheme is 
%revealed in \S\ref{sec:oc-translate}.

%\item \Mo\ is efficient and reliable, thanks to academic and industrial support for the programming languages.
%Our evaluation shows that programs obfuscated by our \mo\ tool are generally faster than those protected by a popular commercial \vbo\ tool.
%Moreover, as the portion of obfuscated code increases, \mo\ stays reliable by producing correct programs
%while the \vbo\ tool has a chance to make protected programs crash or give incorrect output.
\item \Mo\ is not just a single obfuscation algorithm but a general framework. Although we
particularly utilizes Prolog in this paper, there are other languages that can be misused for \mo. For example, 
the New Jersey implementation of ML (SML/NJ)~\cite{sml-nj} does
not even include a runtime stack. Instead, it allocates all frames and closures
on a garbage-collected heap, potentially making 
program analysis much more difficult.
Another example is Haskell, a pure functional language featuring lazy 
evaluation~\cite{Launchbury:1993:NSL:158511.158618}
which can be implemented
with a unique execution model that greatly differs from the traditional
imperative computation~\cite{marlow_2007}.
%\item \Mo\ is open design. Unlike \vbo\ whose implementation is usually kept secret,
%many programming languages have open-source implementations,
%which does not necessarily hurt the security strength of \mo.
%Open design has long been considered as one of the fundamental
%principles in designing secure systems~\cite{1451869}.
%\item \Mo\ is complementary to existing obfuscation techniques
%and can be applied simultaneously with them for even stronger 
%software protection.
\end{itemize}
All these benefits make us believe that \mo\ could be a new direction in software protection.

\section{Misusing Prolog for Obfuscation}\label{sec:background}
In this section we briefly introduce the Prolog programming language
and explain why we can misuse its language features for obfuscation.
\subsection{Prolog Basics}
The basic building blocks of Prolog are \emph{terms}. Both a
Prolog program itself and the data it manipulates are built from
terms. There are three kinds of terms: constants, variables, and
structures. A constant is either a number (integer or real) or
an atom. 
An atom is a general-purpose name, which is similar to a constant string entity in other languages. 
A structure term is of the form $f(t_1, \cdots, t_n)$, where $f$ is a symbol called a \emph{functor} and
$t_1, \cdots, t_n$ are subterms. The number of subterms a functor takes is called its \emph{arity}.
It is allowed to use a symbol with different arities, so the notation `\texttt{f/n}' is used when 
referring to a structure term $f$ with $n$ subterms.

Structure terms become \emph{clauses} when assigned semantics.
A clause can be a fact, a rule, or a query.
A predefined clause is a fact if it has an empty body, 
otherwise it is a rule. For example,
``\texttt{parent(jack,bill).}'' is a fact, which could mean that
``jack is a parent of bill.'' One the other hand, a rule can be like the following line of code:
\begin{lstlisting}[language=Prolog]
grandparent(G,C):-parent(G,P),parent(P,C).
\end{lstlisting}
This rule can be written as the following formula in the first-order logic:
{\small
\[
\begin{array}{r@{~~}l}
  \forall G, C, P. \mathrm{grandparent}(G,C)\leftarrow\mathrm{parent}(G,P) \land \mathrm{parent}(P,C)
\end{array}
\]
}%
Clauses with the same name and the same number of arguments define a relation, namely a \emph{predicate}.
With facts and rules defined, programmers can issue \emph{queries}, which are formulas for the Prolog
resolution system to solve. In accordance with our previous examples, a query could be \texttt{grandparent(G,bill)}
which is basically asking ``who are bill's grandparents?'' 

A Prolog program is a set of terms.
The Prolog resolution engine maintains an internal database of terms throughout program execution,
trying to resolve queries with facts and rules by logical inference. Essentially, computation in
Prolog is reduced to a searching problem. This is different from the commonly seen Turing machine
computation model but the theoretical foundation of logic programming guarantees that Prolog is
Turing complete~\cite{tarnlund1977horn}.

\subsection{Obfuscation-Contributing Features}\label{sec:oc-features}
\ignore{There are numerous language features of Prolog that provide obfuscation effects, e.g.,
unification, backtracking, term indexing, tabling, and multi-stack, etc.
In \mo\ we elaborate the two most salient ones: \textit{unification} and \textit{backtracking}.
Conceptually these two features are part of the foundations of logic programming.}

\subsubsection{Unification}
One of the core concepts in automated logic resolution, hence in logic programming, is unification.
Essentially it is a pattern-matching technique.
Two first-order terms $t_1$ and $t_2$ can be unified if there exists
a substitution $\sigma$ making them identical, i.e., $t_1^\sigma=t_2^\sigma$.
For example, \ignore{a logical variable $X$ can be unified with a constant $c$;}
two terms $k(s(g),Y)$ and $k(X,t(k))$ are unified when
$X$ is substituted by $s(g)$ and $Y$ is substituted by $t(k)$.

Unification is one of the basises of Prolog's computation model.
We show this by example.
The following clause defines a simple ``increment-by-one'' procedure:
\begin{lstlisting}[language=Prolog]
inc(Input,Output):-Output is Input+1.
\end{lstlisting}
Now for a query \texttt{inc(1,R)}, the Prolog resolution engine will first try to unify \texttt{inc(1,R)} with \texttt{inc(Input,Output)},
which means \texttt{Input} should be unified with \texttt{1} and \texttt{Output} should be unified with \texttt{R}.
Once this unification succeeds, the original query is reduced to a subgoal \texttt{Output is Input+1}. Since \texttt{Input} is now unified
with \texttt{1}, \texttt{Input+1} is evaluated as \texttt{2}. 
Finally \texttt{Output} gets unified with \texttt{2} (\texttt{is/2} is the evaluate-and-unify operator predicate),
making \texttt{R} unified with \texttt{2} as well.

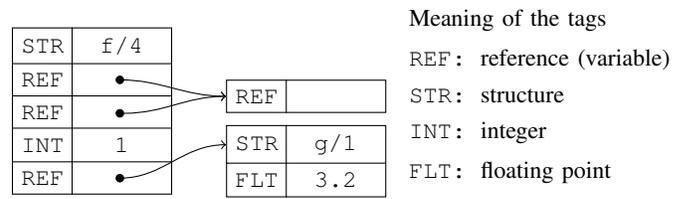
\begin{figure}[t]
  \begin{tikzpicture}
    \tt
    \node[rectangle split, rectangle split parts=5, 
      draw, minimum width=6em,font=\small,
      rectangle split part align={left}] (t1)
         {             
           STR\hspace*{1.5em}f/4
           \nodepart{two}
           REF
           \nodepart{three}
           REF
           \nodepart{four}
           INT\hspace*{2.1em}1
           \nodepart{five}
           REF
         };
         \coordinate[left=.75em of t1.north] (north1);
         \coordinate[left=.75em of t1.south] (south1);
    \draw (north1) -- (south1);
    \coordinate[right=1.05em of t1.north, yshift=-1.98em] (ref1start);
    \node[draw,circle,inner sep=1pt,fill] at (ref1start) {};
    \node[draw,circle,inner sep=1pt,fill,below=1.1em of ref1start] (ref3start) {};

    \coordinate[right=1.05em of t1.south, yshift=.62em] (ref2start);
    \node[draw,circle,inner sep=1pt,fill] at (ref2start) {};

    \node[rectangle split, rectangle split parts=1, 
      draw, minimum width=6em, font=\small, 
      right=4em of ref1start, yshift=-.62em,
      rectangle split part align={left}] (t2)
         { REF\hspace*{4em} };
         \coordinate[left=.75em of t2.north] (north2);
         \coordinate[left=.75em of t2.south] (south2);
    \draw (north2) -- (south2);
    \draw[->] (ref1start) to [out=0,in=180] (t2.west);
    \draw[->] (ref3start) to [out=0,in=180] (t2.west);

    \node[rectangle split, rectangle split parts=2, 
      draw, minimum width=6em, font=\small, 
      right=4em of ref2start, yshift=.64em,
      rectangle split part align={left}] (t3)
         { 
           STR\hspace*{1.5em}g/1
           \nodepart{two}
           FLT\hspace*{1.5em}3.2
         };
         \coordinate[left=.75em of t3.north] (north3);
         \coordinate[left=.75em of t3.south] (south3);
    \draw (north3) -- (south3);
    \coordinate[yshift=.62em] (ref2end) at (t3.west);
    \draw[->] (ref2start) to [out=0,in=180] (ref2end);

    \node[rectangle split, rectangle split parts=5, 
      font=\small,
      right=.5em of t2,
      rectangle split part align={left}] (t4)
         {             
           \textrm{Meaning of the tags}
           \nodepart{two}
           REF: \textrm{reference (variable)}
           \nodepart{three}
           STR: \textrm{structure}
           \nodepart{four}
           INT: \textrm{integer}
           \nodepart{five}
           FLT: \textrm{floating point}
         };
    
  \end{tikzpicture}
  \caption{An example memory representation of term \texttt{f(X,Y,1,g(3.2))} in Prolog, where both \texttt{X} and \texttt{Y} are unified with another variable which itself is un-unified.}
  \label{fig:uni-imp}
\end{figure}

To support unification, Prolog implements terms as vertices in directed acyclic graphs.
Each term is represented by a \texttt{<tag,content>} tuple, 
where \texttt{tag} indicates whether the type of the term
and \texttt{content} is either the value of a constant or the address of
the term the variable is unified with.
\F~\ref{fig:uni-imp} is an example showing how Prolog may represent a term in memory~\cite{ait-kaci_warrens_1991}.

%Conceptually, the unification system is in the core of automated logic resolution
%and can be seen as the foundation of most unique language features of Prolog.
%Moreover, unification is an obfuscation-contributing factor on its own.
Unification makes data shapes in Prolog program memory dramatically different from C and much more obscure. 
The graph-like implementation of unification 
poses great challenges to binary data shape analyses which
aim to recover high-level data structures from binary program 
images~\cite{Jones:1982:FAI:582153.582161,Ghiya:1996:TDC:237721.237724,Sagiv:1998:SSP:271510.271517,Cozzie:2008:DDS:1855741.1855759}. 
Even if some of the graph structures can be identified,
there is still a gap between this low-level representation and the logical organization of original data,
which harshly tests attackers' reverse engineering abilities.
Unification also complicates data access. 
To retrieve the true value of a variable, the Prolog engine has to iterate the entire 
unification list. It is well known that static analysis is weak against loops and indirect memory access.
Also, the tags in the term tuples are encoded as bit fields, meaning that
bit-level analysis algorithms are required to reveal the semantics of a binary compiled from Prolog code.
However, achieving bit-level precision is another great technical challenge for both static and
dynamic program analyses, mainly because of scalability issues~\cite{sepp2011,Kettle:2008:BRA:1512464.1512474,Drewry:2007:FEA:1323276.1323277,yadegari2014}.

\subsubsection{Backtracking}
\begin{figure*}[!t]
  \centering
  \begin{minipage}[b][][t]{.45\textwidth}
    \begin{minipage}[b][3.2cm][t]{.46\textwidth}
      \begin{lstlisting}[basicstyle=\ttfamily\footnotesize]
int foo(int sel, 
        int x, int y)
{
  int ret;
  if(sel==1) 
    ret=x;
  else
    ret=y;
  return ret;
}
      \end{lstlisting}
    \end{minipage}%\hfill
    \begin{minipage}[b]{.53\textwidth}
      \centering
      \scalebox{.65}{
        \begin{tikzpicture}[node distance=1.9cm]
          \tt
          \tikzstyle{block} = [rectangle, draw, text width=6.7em, text centered]
          \tikzstyle{line} = [draw, -latex',line width=1pt]
          \node [block, text width=9em] (init) {foo(sel,x,y)};
          \node [block, below of=init, yshift=.5cm, text width=8em] (if) {if(sel == 1)};
          \node [block, below left of=if, xshift=-1em] (true) {ret = x};
          \node [block, below right of=if,xshift=1em] (false) {ret = y};
          \node [block, below left of=false,xshift=-1em] (ret) {return ret};
          \path [line] (init) -- (if);
          \path [line] (if) -- (true) node [near end, above, fill=white, inner sep=0em, yshift=.3em, xshift=-.6em] {\sf true};
          \path [line] (if) -- (false) node [near end, above, fill=white, inner sep=0em, yshift=.3em, xshift=.6em] {\sf false};
          \path [line] (true) -- (ret);
          \path [line] (false) -- (ret);
          \node [right of=ret, xshift=.5cm] {}; % this node is to increase white space at the right side of the picture. A dirty way to do it, but works.
        \end{tikzpicture}
      }
    \end{minipage}%
  \end{minipage}\vrule\hfill%
  \begin{minipage}[b][][t]{.53\textwidth}
    \begin{minipage}[b][2.5cm][t]{.25\textwidth}
      \begin{lstlisting}[basicstyle=\ttfamily\footnotesize]
pfoo(Sel,X,Y,R) :-
  (Sel =:= 1 ->
     R is X);
  (R is Y).
      \end{lstlisting}
    \end{minipage}\hfill%
    \begin{minipage}[b]{.75\textwidth}
      \centering
      \scalebox{.65}{
        \begin{tikzpicture}[node distance=1.9cm]
          \tt
          \tikzstyle{block} = [rectangle, draw, text width=6.7em, text centered,]
          \tikzstyle{line} = [draw, -latex',line width=1pt]
          \node [block, text width=9.8em] (init) {pfoo(Sel,X,Y,R)};
          \node [block, right = 2.5cm of init] (lt5) {Sel =:= 1};
          \node [block, below of = init,yshift=0cm] (sr1) {{\sf Unification:}\\R is X};
          \node [block, below of = sr1,xshift=-3em] (end) {\sf Next Clause};
          \node [block, below of =lt5 ,yshift=0cm, text width=7em] (re) {\sf Resolution\\Failure Handler};
          \node [block, right = of end, xshift=0em] (sr2) {{\sf Unification:}\\R is Y};
          \node [block, right = of sr2,xshift=-2.5em,text width=5em] (fail) {\sf Fail};

          \path [line] (init) -- (lt5);
          \path [line] (lt5) to node [fill=white, inner sep=0,above,yshift=1ex] {\sf true} (sr1);
          \path [line,arrows=->>] (sr1) to node [fill=white, inner sep=0,above,yshift=1ex,text=black] {\sf fail} (re);
          \path [line] (lt5) to node [fill=white, inner sep=0,above,text=black,xshift=1.2em] {\sf false} (re);
          \path [line, dashed, arrows=->>] (re) to [bend left=20] node {} (sr2);
          \path [line, dashed] (re) to node {} (fail);
          \path [line, dashed] (sr1) to node [fill=white, inner sep=0,above,] {\sf succeed} (end);
          \path [line, dashed] (sr2) to node [fill=white, inner sep=0,above,yshift=.5ex] {\sf succeed} (end);
          \path [line,arrows=->>] (sr2) to [bend left=20] node [fill=white, inner sep=0,xshift=-1em,text=black] {\sf fail} (re);
        \end{tikzpicture}
      }
    \end{minipage}%
  \end{minipage}%
  \caption{Different control flows of C and Prolog binaries implementing the same algorithm, due to different execution models.
In the Prolog graph, dashed lines indicate indirect jumps and arrows with the same pattern indicate feasible paths through the resolution failure handler. Both control flow graphs are summarized from post-compilation binaries.}\label{fig:example}
\end{figure*}
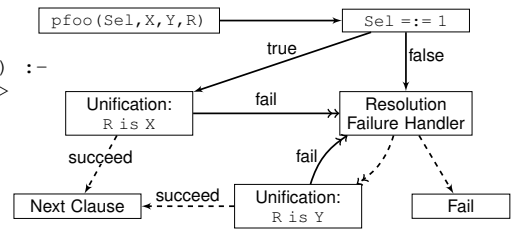

Different from Prolog unification which mainly obfuscates program data,
the backtracking feature obfuscates the control flow.
Backtracking is part of the resolution mechanism in Prolog.
As explained earlier, finding a solution for a resolvable formula is essentially
searching for a proper unifier, namely a substitution,
so that the substituted formula can be expanded to consist of only facts
and other formulas known to be true. 
Since there may be more than one solution for a unification problem instance, 
it is possible that the resolution process will unify two terms
in the way that it makes resolving the formula later unfeasible. 
As a consequence,
Prolog needs a mechanism to roll back from an incorrect proof path, which is called
backtracking.

To make backtracking possible, Prolog saves the program state before taking one
of the search branches. This saved state is called a ``choice-point'' by Prolog
and is similar to the concept of ``continuation'' in functional programming.
When searching along one path fails, the resolution engine will restore the 
latest choice-point and continue to search through one of the untried branches.

This \emph{search-and-backtrack} execution model leads to a totally different control flow scheme in
Prolog programs at the low level, compared to programs in the same logic written by C.
\F~\ref{fig:example} is an example where a C function is transformed into a Prolog clause by our tool 
\babel\ (with manual edits to make the code more readable), along with
the program execution flows before and after \babel\ transformation. 
The real control flow of the Prolog version of the function is much more complicated than presented,
and we have greatly simplified the flow chart for readability.
In the Prolog part of \F~\ref{fig:example}, a choice point is created
right after the execution flow enters the predicate \texttt{pfoo}
which is a disjunction of two subclauses. 
The Prolog resolution routine will first try to satisfy the first subclause.
If it fails, the engine will backtrack to the last choice-point
and try the second subclause.

Due to the complicated backtracking model, a large portion of control flow transfers in
Prolog are indirect.
The implementation of backtracking also involves techniques such as
long jump and stack unwinding. 
Clearly, Prolog has a much more obscure low-level execution model compared to C, and
imperative programming in general, from
the perspective of static analysis.
Different from some other control flow obfuscation techniques that inject fake control flows 
which are never feasible at run time,
Prolog's backtracking actually happens during program execution,
making \mo\ also resilient to dynamic analysis.
Most importantly, after the C-to-Prolog translation
the original C control flows are reformed with a completely different programming paradigm,
which is fundamentally different from existing control-flow based obfuscation techniques.

\section{Technical Challenges}
\label{sec:cpob}
To make use of Prolog's execution model for obfuscating C programs,
we need a translation technique to forge the Prolog counterpart of a C function.
At this point, there are various challenges to resolve. 

\subsection{Control Flow}
As an imperative programming language,
C provides much flexibility of crafting program control 
flows almost with language key words such as continue, break, and return. 
Prolog programs, however, have to follow the general evaluation procedure of logical 
formulas, which inherently forbids some ``fancy'' control flows allowed by C. 

\subsection{Memory Model}
In C programming, many low-level details are not opaque to programmers.
As for memory manipulation, C programmers can access almost arbitrary memory locations via pointers.
Prolog lacks the semantics to express direct memory access. 
Moreover, the C memory model is closely coupled with other sub-structures of the language, 
e.g., the type system. C types are not only logical abstractions but are also implications on
low-level memory layouts of the data.
For instance, logically adjacent elements in a C array and fields in a C struct are 
also physically adjacent in memory. 
Therefore, some logical operations on C data structures can be implemented as direct memory 
accesses which are semantically equivalent only with the C memory modeling. 
Below is an example.
\begin{lstlisting}
struct ty { 
 int a;
 int b; 
} s[2];

/* Equivalent to s[0].a=s[0].b=s[1].a=0;
 * with many compilers and architectures */
memset((void*)s, 0, 3*sizeof(int));
\end{lstlisting}
Translating the code snippet above into pure Prolog could be difficult
because the translator will have to infer the logic effects of the \texttt{memset} statement. 

\subsection{Type Casting}
C type casting is of full flexibility in the sense that a C programmer can cast any type to any other type,
no matter the conversion makes sense or not.
This can be realized by violating the \textit{load-store consistency}, 
namely storing a variable of some type into a memory location and later loading the content of the same chunk of memory into a variable of another type. 
The C union type is a high-level support for type castings that breaks the load-store consistency,
but C programmers can choose to use pointers to directly achieve the same effect.
Imitating this type casting system could be a notable challenge for other languages.

\section{C-to-Prolog Translation}\label{sec:design}
This section explains how we address the challenges mentioned in the
previous section.
\longversion{Considering the many obstacles for developing a complete translation from 
C to Prolog, we do not seek to obtain a pure Prolog version 
of the original C program.
The section explains how we address the challenges mentioned in the
previous section and how we develop a partial C-to-Prolog translation method,
which is suitable for \mo.}

\subsection{Control Flow Regularization}\label{sec:cfr}
There has been a large amount of research on refining C program control flows,
especially on eliminating goto statements~\cite{knuth1971notes,ramshaw1988eliminating,williams1985restructuring}.
For now, we consider that goto elimination is a solved problem and assume
the C programs to be protected do not contain goto statements.
Given a C function without goto statements,
there are two control flow patterns that cannot be directly
adopted by Prolog programming, i.e., control flow cuts and loops. 
We call these patterns \emph{irregular} control flows.

\subsubsection{Control Flow Cuts}
Control flow cuts refer to the termination of control flows in the middle of a C function,
for example:
\begin{lstlisting}
int foo (int m, int n) {
 if(m) 
  return n; // Flow of if branch ends
 else
  n=n+1;
 n=n+2;
 return n;  // Flow of else branch ends
}
\end{lstlisting}

The C language grants programmers much freedom in building control flows,
even without using goto statements.
In Prolog, however, control flows have to be routed based on the short-circuit rules in evaluating logical expressions.
With short-circuit effects,
parallel statements can be connected by disjunction and sequential statements
can be connected by conjunction.
To show why the control flow pattern in the C code above cannot be implemented by only adopting short-circuit rules,
consider a C function with body \mbox{\texttt{\{if(e) \{a;\} else \{b;\} c;\}}}.
Naturally, it should be translated into a Prolog sentence \texttt{(((e->a);b),c)},
where \texttt{->} denotes implication, 
\texttt{;} denotes disjunction, 
and \texttt{,} denotes conjunction.
However, this translation is not semantics-preserving when \texttt{a} is a return statement,
because if the clause \texttt{a} is evaluated, at least one of \texttt{b} and \texttt{c} has to be evaluated 
to decide the truth of the whole logic formula.

We fix control flow cuts by
replicating and/or reordering basic blocks syntactically subsequent to the cuts.
For example, we rewrite the previously shown C function into the following structure:
\begin{lstlisting}
int foo (int m, int n) {
 if(m)
  return n;
 else {
  n=n+1;
  { n=n+2; return n; }
 }
}
\end{lstlisting}
After the revision, the C code is naturally translated into a new Prolog clause
\texttt{(e->a);(b,c)}, which is consistent with the original C semantics.

\subsubsection{Loops}
Most loops cannot be directly implemented in Prolog.
The fundamental reason is that Prolog does not allow
unifying a variable more than once.
We can address this problem by transforming loops into recursive functions,
but in-loop irregular control flows complicate the situation.
The irregularity comes from the use of keywords ``continue'' and ``return.'' 
A continue statement cuts the control flow in the middle of a loop, 
bringing up a problem similar to the aforementioned asymmetric returns in functions.
As such, irregular control flows resulting from continue statements can be regularized in the same way,
i.e., replicating and/or reordering basic blocks syntactically subsequent to continue statements.

Like a continue statement, a return statement also cuts the control flow in a loop, but its impact
reaches outside because it cuts the control flow of the function enclosing the loop.
Hence, a recursive Prolog predicate transformed from a loop needs an extra argument to carry 
a flag indicating whether an in-loop return has occurred.

\subsection{C Memory Model Simulation}
As stated in \S\ref{sec:cpob}, the C memory model is closely coupled
with other parts of the language and it is hard to separate them.
However, \mo\ keeps the original C memory model,
making preserving semantic equivalence much easier.
In our design, the Prolog runtime is embedded in the C execution environment,
so it is possible for Prolog code to directly operate memories within a program's address space. 

The way we handle C memory simulation illustrates the advantage of developing language 
translations for obfuscation purposes.
Unlike tools seeking complete translation from C to other languages,
\mo\ does not have to mimic C memory completely with target language features (e.g., converting C pointers to Java references~\cite{demaine1998c}),
meaning we can reduce translation complexity and circumvent various limitations.
That being said,
partially imitating the C memory model in Prolog is still a non-trivial task.

\subsubsection{Supporting C Memory-Access Operators}
The first step to simulating the C memory model is to support pointer operations.
We introduce the following new clauses into our target Prolog language:
\begin{lstlisting}[language=Prolog]
rdPtrInt(+Ptr, +Size, -Content)
wrPtrInt(+Ptr, +Size, +Content)
rdPtrFloat(+Ptr, +Size, -Content)
wrPtrFloat(+Ptr, +Size, +Content)
\end{lstlisting}
These clauses are implemented in C. \texttt{rdPtrInt/3} and \texttt{rdPtrFloat/3} allow us to load 
the content of a memory cell (address and size indicated by \texttt{Ptr} and \texttt{Size}, respectively) into a Prolog variable \texttt{Content}.
Similarly, \texttt{wrPtrInt/3} and \texttt{wrPtrFloat/3} can write the content of a Prolog variable 
into a memory cell.
These four clauses simulate the behaviors of the ``pointer dereference'' operator (\texttt{*}) in C.

In addition to read-from-pointer and write-to-pointer operations, C also
has the ``address-of'' operator which takes an lvalue, i.e., an expression that is allocated a
storage location, as the operand and returns
its associated storage location, namely address. There is no need to explicitly support this
operator in Prolog because the address of any lvalue in C has a static representation
which is known by the compiler.\footnote{For example, a local variable is usually
allocated on the stack and the compiler will have a static expression of its address.
On x86, the expression is likely to be \texttt{\$offset(\%ebp)} or \texttt{\$offset(\%esp)},
where \texttt{\$offset} is a constant. Compilers can also decide how to statically represent
the addresses of global variables.}
We can obtain the results of ``address-of'' operations 
in the C environment and pass those values into the Prolog environment as arguments.

We also handle several C syntax sugers related to memory access:
``subscript'' (\texttt{[]}) and ``field-of'' (\texttt{.} and \texttt{->}).
We convert these operators into equivalent combinations of pointer arithmetic and dereference
so that we do not need to coin their counterparts in Prolog. 
This conversion requires assumptions on compiler implementation and target architecture to calculate type sizes and
field displacements. 

\subsubsection{Maintaining Consistency}
It is a natural scheme that a C-to-Prolog translation maps every C 
variable to a corresponding Prolog variable. 
Prolog does not allow variable update,
but we can overcome this restriction by transforming C code into a form close to 
static single assignment (SSA), 
in which variables are only initialized at one program location and never updated.
In the strict SSA form, variables can only be statically initialized once 
even if the scopes are disjoint.
Prolog does not require this because the language checks re-unification at run time, 
meaning variables can be updated in exclusively executed parts of the program,
e.g., the ``then'' and ``else'' branches of the same if statement.
Therefore, we do not need to implement the $\phi$ function in our SSA transformation.

The SSA transformation can be implemented by renaming variables.
The challenging part is that simply renaming variables 
in the original C code could break program semantics because 
of side effects caused by memory operations, i.e., variable contents can 
be accessed without referring to variable names.
This is the consistency problem we have discussed earlier.
\F~\ref{fig:side-effects} shows an instance of the problem.

\newsavebox{\firstlisting}
\begin{lrbox}{\firstlisting}% Store first listing
\begin{minipage}{0.47\linewidth}
\begin{lstlisting}[basicstyle=\ttfamily\footnotesize]
a=0;
p=&a;// p points to a
a=1; // a gets 1
b=*p;// b gets a(1)

c=0;
p=&c;
c=1; // c gets 1
*p=3;// c gets 3
d=c; // d gets c(3)
\end{lstlisting}
\end{minipage}
\end{lrbox}
\newsavebox{\secondlisting}
\begin{lrbox}{\secondlisting}% Store second listing
\begin{minipage}{0.51\linewidth}
\begin{lstlisting}[frame=leftline,basicstyle=\ttfamily\footnotesize]
a1=0;
p1=&a1;// p1 points to a1
a2=1;  // a2 gets 1
b1=*p1;// b1 gets a1(0)

c1=0;
p2=&c1;// p2 points to c1
c2=1;  // c2 gets 1
*p2=3; // c1 gets 3
d1=c2; // d1 gets c2(1)
\end{lstlisting}
\end{minipage}
\end{lrbox}
\newsavebox{\thirdlisting}
\begin{lrbox}{\thirdlisting}% Store third listing
\begin{minipage}{0.13\linewidth}
\begin{lstlisting}[basicstyle=\ttfamily\footnotesize]



a=0;     
p=&a;
a=1;

b=*p;

c=0;
p=&c;
c=1;
*p=3;

d=c;

\end{lstlisting}
\end{minipage}%
\end{lrbox}
\newsavebox{\forthlisting}
\begin{lrbox}{\forthlisting}% Store forth listing
\begin{minipage}{0.32\linewidth}
\begin{lstlisting}[frame=leftline,basicstyle=\ttfamily\footnotesize]
pa=&a;
pc=&c;

a=0;
p=&a;
a=1;
*pa=a;// Flush
b=*p;

c=0;
p=&c;
c=1;
*p=3;
c=*pc;// Reload
d=c;
\end{lstlisting}
\end{minipage}%
\end{lrbox}
\newsavebox{\fifthlisting}
\begin{lrbox}{\fifthlisting}% Store fifth listing
\begin{minipage}{0.50\linewidth}
\begin{lstlisting}[frame=leftline,basicstyle=\ttfamily\footnotesize]
pa=&a1;// const pointer
pc=&c1;// const pointer

a1=0;
p1=&a1;
a2=1;  // a2 gets 1
*pa=a2;// a1 gets a2(1)
b1=*p1;// b1 gets a1(1)

c1=0;
p2=&c1;
c2=1;
*p2=3; // c1 gets 3
c3=*pc;// c3 gets c1(3)
d1=c3; // d1 gets c3(3)
\end{lstlisting}
\end{minipage}%
\end{lrbox}
\begin{figure}[!t]
  \vspace*{-12pt}
  \subfloat[\scriptsize Original]{\usebox{\firstlisting}}\label{fig:side-effects-orign}\hfill%
  \subfloat[\scriptsize Renamed]{\usebox{\secondlisting}}\label{fig:side-effects-after}
  \caption{Memory operations affecting the correctness of C source code SSA renaming.}\label{fig:side-effects}
%\end{figure}%
%\begin{figure}[!t]
%  \vspace{-.2in}
  \subfloat[\scriptsize Orig.]{\usebox{\thirdlisting}}\label{fig:flush-reload-origin}\hfill%
  \subfloat[\scriptsize With flush and reload]{\usebox{\forthlisting}}\label{fig:flush-reload-insert}\hfill%
  \subfloat[\scriptsize Renamed with flush and reload]{\usebox{\fifthlisting}}\label{fig:flush-reload-ssa}
  \caption{Semantic-preserving SSA renaming on C source code with the presence of pointer operations.}
  \label{fig:flush-reload}
\end{figure}

To address this issue,
we keep the addresses of local variables and parameters if they are possibly accessed via pointers.
Then we flush variable contents back to the memory before a read-from-pointer operation and reload
variable contents from the memory after a write-to-pointer operation.
Inter-procedural pointer dereferences are also taken into account.
When callee functions accept pointers as arguments,
we do variable flush before function calls and do variable reload after.
The flush makes sure that changes made by Prolog code are committed to the underlying C memory before they are read again.
Similarly, the reload assures that values unified with Prolog logical variables are always consistent with the content in C memory.
%The flush and reload method is a basic technique in maintaining cache coherence for multi-core microprocessors~\cite{patterson2013computer}.
%
We perform a sound points-to analysis to compute the set of variables that need to be reloaded or flushed at each program point.
After inserting the flush and reload operations, 
the SSA variable renaming no longer breaks the original program semantics.
%Moreover, with the SSA transformation applied a simple one-to-one
%mapping from C variables to Prolog variables is semantic preserving.
\F~\ref{fig:flush-reload} illustrates our solution based on the example in \F~\ref{fig:side-effects}.

\subsection{Supporting Other C Features}
%\textit{Struct, union, and array.} 
\subsubsection{Struct, Union, and Array} 
In \S\ref{sec:cpob}, we showed that C data types 
like struct and array can be manipulated via memory access.
Since we have already built support for the C memory model in Prolog, 
the original challenge now becomes a shortcut to supporting C struct, union, and array.
We simply transform the original C code and implement all 
operations on structs, unions, and arrays through pointers. 
After this transformation the primitive data types provided by Prolog are enough to represent any C data structure.

%\textit{Type casting.} 
\subsubsection{Type Casting} 
With our C memory simulation method, 
supporting type castings performed via pointers does not require additional effort, 
even if they may violate the load-store consistency.
As for explicit castings, e.g., from integers to floating points, we utilize the
built-in Prolog type casting clauses like \texttt{float/1}.

%\textit{External and indirect function call.}
\subsubsection{External and Indirect Function Call}
Since the source code of library functions is usually unavailable, 
translating them into Prolog is not an option. 
In general, translations of \mo{} can support external subroutine invocation with the help of foreign language interfaces.
As for C+Prolog obfuscation, most Prolog implementations provide the interface for calling C functions from a Prolog context.
The same interface can also be used to invoke functions via pointers.

\subsection{Obfuscating Translation}
\label{sec:oc-translate}
Our translation scheme fully exploits the obfuscation-contributing features introduced 
in \S\ref{sec:oc-features}, generally because:
\begin{itemize}
\item The conversion from C data structures to Prolog data structures happens by default,
and every C assignment is translated to Prolog unification.
\item Intra-procedural control-flow transfers originally coded in C are now implemented by Prolog's backtracking mechanism.
This significantly complicates the low-level logic of the resulting binaries.
\end{itemize}

Especially, we would like to highlight the method we use to support the C memory model in Prolog.
At the high level, the original C memory layout is kept after the translation. However,
the behavior of the C-part memory becomes much different from the original program.
To maintain the consistency between the C-side memory and Prolog-side memory, we introduce
the flush-reload method which disturbs the sequence of memory access.
In this way, the memory footprint of the obfuscated program is no longer what it was during
program execution.

We believe our translation method is one of the factors that make \mo\ resilient to
both semantics-based and syntax-based binary diffing, as will be shown in \S\ref{sec:resilience}.

\section{Implementation of \babel}
\label{sec:impl}
\ignore{In this section, we present the implementation of \babel, a prototype of \mo.}

%\subsection{Overview}
\babel\ is our \mo\ prototype.
The workflow of \babel\ has three steps: 
C code preprocessing, 
C-to-Prolog translation, and C+Prolog compilation.
The preprocessing step reforms the original C source code 
so that the processed program becomes suitable for line-by-line translation to Prolog. 
The second step translates C functions to Prolog predicates.
In the last step, \babel\ combines C and Prolog code together with a carefully designed interface.

We choose GNU Prolog~\cite{diaz2001} as the Prolog implementation to employ in \babel.
Like many other Prolog systems, GNU Prolog compiles Prolog source into the ``standard'' Warren Abstract Machine (WAM)~\cite{warren1983} instructions.
What is desirable to us is that GNU Prolog can further compile WAM code into native code. 
This feature makes \babel\ more distinguishable from virtualization-based obfuscation tools
which compile the original program to bytecode and execute it with a custom virtual machine.
%We have a more detailed discussion on the differences between \mo\ and \vbo\ in \S\ref{sec:discussion}.

\subsection{Preprocessing and Translating C to Prolog}
Before actually translating C to Prolog, we need to preprocess the C code first.
The preprocessing includes the following steps, which is done with the help of the CIL library~\cite{CIL}.
\begin{enumerate}%[nosep]
\item Simplify C code into the three-address form without switch statements and ternary conditional expressions.
\item Convert loops to tail-recursive functions.
\item Eliminate control flow cuts. %Regularize conditional branches with asymmetric returns.
\item Transform operations on global, struct, union, and array variables into pointer operations.
\item Perform variable flush and reload whenever necessary. %at appropriate program points.
\item Eliminate all memory operators except pointer dereferences. % i.e., ``address-of'', ``subscript'', and ``field-of'' operations.
\item Rename variables so that the C code is in a form close to SSA.
\end{enumerate}

After preprocessing, we can translate C to Prolog line by line. 
The translation rules are listed in \F~\ref{fig:trans-rules}. 
Note that by the time we start translating C to Prolog, 
the preprocessed C code does not contain any switch and loop statements, 
because they are transformed into either nested if statements or recursive functions. 
As discussed in \S\ref{sec:cfr}, we do not consider goto statements.

We take translating arithmetic and logical expressions as a trivial task, 
but that leads to a limitation in our translation.
Due to the fact that Prolog does not subdivide integer types,
integer arithmetics in Prolog are not equivalent to their C counterparts.
For example, given two C variables \texttt{x} and \texttt{y} of type \texttt{int} (4 bytes long) and
their addition \texttt{x+y}, the equivalent expression in Prolog should be
\texttt{(X+Y){/\symbol{92}}0xffffffff}, assuming that \texttt{X} and \texttt{Y} are the corresponding logical variables of \texttt{x} and \texttt{y}. 
Therefore, if a C program intentionally relies on integer overflows or underflows,
there is a chance that our translation will fail.
However, fully emulating C semantics incurs significant performance penalty.

Previous work on translating C to other languages faces the same issue, 
and many of them chose to ignore it~\cite{Buddrus:1998:CMC:330560.331015, martin_strategies_2001,trudel_c_2012}.
The C-to-JavaScript converter Emscripten provides the option to fully emulate the C semantics~\cite{zakai_emscripten:_2011}.
It also has a set of optional heuristics to infer program points
where full emulation is necessary, but that method is not guaranteed to work correctly.
We do not particularly take this issue into account when implementing \babel. 
However, we expect \babel's translation to have a low failure chance thanks to the employment
of write-to-pointer operations in Prolog and the variable flush-reload method. 
Since the write-to-pointer operation specifies data sizes, 
the truncation automatically takes place whenever an integer variable is flushed and reloaded.
In GNU Prolog on 64-bit platforms,
all integers are represented by 61-bit two's complement (3 bits are occupied by a WAM tag),
which is large enough to hold most practical integer and
pointer\footnote{Most 64-bit CPUs only implement a 48-bit virtual address space.} values.

\begin{figure*}[!ht]
  \centering
\resizebox{\textwidth}{!}{
\tt
  \begin{tabular}{llll}
    \rm{Assignment} &  (foo = $e$;)$^\mathcal{T}$ & $\rightarrow$ & (Pfoo is $e^\mathcal{T}$) \\
    \rm{Pointer arithmetic}  & (p2 = p1 + intVal;)$^\mathcal{T}$ | $\mathrm{TypeOf}$(p1) = $T$* & $\rightarrow$ &  ($\sigma(\mbox{p2})$ is $\sigma(\mbox{p1})$ + $\mathrm{SizeOf}$($T$) * $\sigma(\mbox{intVal})$)\\
    \rm{Pointer dereference} & (foo = *p;)$^\mathcal{T}$ | $\mathrm{TypeOf}$(p) = $T$* & $\rightarrow$ & rdPtr($\sigma(\mbox{foo})$, $\mathrm{\mathrm{SizeOf}}$($T$), $\sigma(\mbox{p})$) \\
    \rm{Write by pointer} & (*p = foo;)$^\mathcal{T}$ | $\mathrm{TypeOf}$(p) = $T$* & $\rightarrow$ & wrPtr($\sigma(\mbox{p})$, $\mathrm{SizeOf}$($T$), $\sigma(\mbox{foo})$) \\
    \rm{Empty Block} & (\{\})$^\mathcal{T}$ & $\rightarrow$ & (true) \\
    \rm{Non-empty Block} & (\{$s_1 \cdots s_n$\})$^\mathcal{T}$ & $\rightarrow$ & ($s_1^\mathcal{T}$, $\cdots$, $s_n^\mathcal{T}$) \\
    \rm{Conditional} &  (if ($e$) \{$b_{then}$\} else \{$b_{else}$\})$^\mathcal{T}$ & $\rightarrow$ & ($e^\mathcal{T}$, \{$b_{then}$\}$^\mathcal{T}$; \{$b_{else}$\}$^\mathcal{T}$) \\
    \rm{Function call} &  (ret = fun(a1, $\cdots$, an);)$^\mathcal{T}$ & $\rightarrow$ & predFun($\sigma(\mbox{a1})$, $\cdots$, $\sigma(\mbox{an})$, $\sigma(\mbox{ret})$)\\
    \rm{Indirect function call} &  (ret = funptr(a1, $\cdots$, an);)$^\mathcal{T}$ & $\rightarrow$ & predIndFun($\sigma(\mbox{funptr})$, $\sigma(\mbox{a1})$, $\cdots$, $\sigma(\mbox{an})$, $\sigma(\mbox{ret})$)\\
    \rm{Function return} &  (return $e$;)$^\mathcal{T}$ & $\rightarrow$ & ($\mathcal{R}$ is $e^\mathcal{T}$)\\
    \rm{Function definition} & (fun($T_1$ a1, $\cdots$, $T_n$ an) \{$b_{body}$\})$^\mathcal{T}$ & $\rightarrow$ & predFun($\sigma(\mbox{a1})$, $\cdots$, $\sigma(\mbox{an})$) :- $b_{body}^\mathcal{T}$.
  \end{tabular}
}
  \caption{Definition of $\mathcal{T}$, \babel's C-to-Prolog translation. $e$, $s$, $b$, and $T$ denote C expressions, statements, blocks, and types, respectively. $\sigma$ is the bijective mapping from C identifiers to corresponding Prolog identifiers. $\mathcal{R}$ denotes the Prolog identifier used to hold the returned value in the translated predicate. \texttt{predFun} can be either a real Prolog predicate or a wrapper of a foreign C function, depending on whether the target function is translated or not. \texttt{predIndFun} is a wrapper for a special foreign C function which further calls into \texttt{funptr} with given arguments.}
  \label{fig:trans-rules}
\end{figure*}

\subsection{Combining C and Prolog}

\shortlong%
{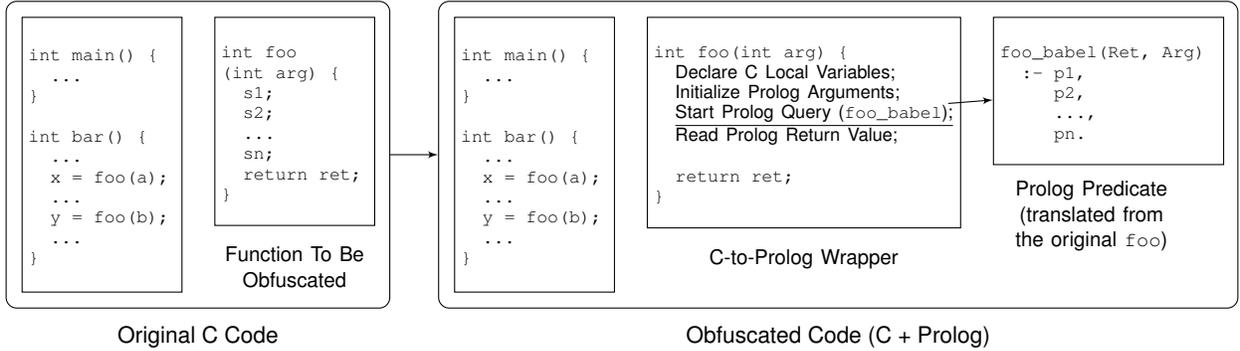
\begin{figure}[!t]
  \centering
  \begin{tikzpicture}[scale=0.65,every node/.style={transform shape}]
    \tikzstyle{wrapper} = [draw, rectangle, rounded corners]
    \tikzstyle{block} = [rectangle, draw, text centered, inner sep=.1cm]
    \tikzstyle{line} = [draw, -latex']
    \node[block,text width=2.3cm] (orig-foo) {
      \begin{lstlisting}[basicstyle=\ttfamily\footnotesize]
int foo
(int arg) {
  s1;
  s2;
  ...
  sn;
  return ret;
}
      \end{lstlisting}

   };
   \node[below=.2cm of orig-foo,text width=2.3cm, text centered] {\small \sf Function To Be\\Obfuscated};
   \node[wrapper,minimum width=3cm,minimum height=4.8cm, below right= -.25cm and -.25cm of orig-foo.north west] (orig-wrapper) {};
   \node[below=.15cm of orig-wrapper] {\sf Original C Code};

   \node[wrapper,minimum width=9.25cm, minimum height=4.8cm, right=.75cm of orig-wrapper] (obf-wrapper) {};
   \path[line] (orig-wrapper) -- (obf-wrapper);
   \node[below=.15cm of obf-wrapper] {\sf Obfuscated Code (C + Prolog)};
   \node[block,text width=4.7cm, below right=.25cm and .25cm of obf-wrapper.north west] (obf-main) {
     \begin{lstlisting}[escapechar=@,basicstyle=\ttfamily\footnotesize]
int foo(int arg) {

  @\it{\textbf{\sf Declare C Local Variables;}}@
  @\it{\textbf{\sf Initialize Prolog Arguments;}}@
  @\it{\underline{\sf Start Prolog Query (\texttt{foo\_babel});}}@
  @\it{\textbf{\sf Read Prolog Return Value;}}@

  return ret;
}
      \end{lstlisting}
   };

   \node[block,text width=3.4cm, below right=0cm and .25cm of obf-main.north east] (obf-pro) {
      \begin{lstlisting}[basicstyle=\ttfamily\footnotesize]
foo_babel(Ret, Arg)
  :- p1,
     p2,
     ...,
     pn.
      \end{lstlisting}
   };

   \node[below=.2cm of obf-main,text width=4.8cm, text centered] (obf-foo-text) {\small \sf C-to-Prolog Wrapper};
   \node[above right=0cm and .25cm of obf-foo-text.east,text width=3.2cm, text centered] {\small \sf Prolog Predicate (translated from the original \texttt{foo})};
   \coordinate[above right=.3cm and -.2cm of obf-main.east] (ctr);
   \path[line] (ctr) -- (obf-pro);
  \end{tikzpicture}
  \caption{The context for executing obfuscated code in \babel.}
  \label{fig:context}
\end{figure}
}%
{\begin{figure*}[!t]
  \centering
  \begin{tikzpicture}[scale=0.85,every node/.style={transform shape}]
    \tikzstyle{wrapper} = [draw, rectangle, rounded corners]
    \tikzstyle{block} = [rectangle, draw, text centered, inner sep=.1cm]
    \tikzstyle{line} = [draw, -latex']
    \node[block,text width=2.3cm, minimum height=4.3cm] (orig-main) {
      \begin{lstlisting}[basicstyle=\ttfamily\footnotesize]
int main() {
  ...
}

int bar() {
  ...
  x = foo(a);
  ...
  y = foo(b);
  ...
}
      \end{lstlisting}
   };

   \node[block,text width=2.3cm,below right=0cm and .5cm of orig-main.north east] (orig-foo) {
      \begin{lstlisting}[basicstyle=\ttfamily\footnotesize]
int foo
(int arg) {
  s1;
  s2;
  ...
  sn;
  return ret;
}
      \end{lstlisting}

   };
   \node[below=.2cm of orig-foo,text width=2.3cm, text centered] {\small \sf Function To Be\\Obfuscated};
   \node[wrapper,minimum width=6.0cm,minimum height=4.8cm, below right= -.25cm and -.25cm of orig-main.north west] (orig-wrapper) {};
   \node[below=.15cm of orig-wrapper] {\sf Original C Code};

   \node[wrapper,minimum width=12.5cm, minimum height=4.8cm, right=.75cm of orig-wrapper] (obf-wrapper) {};
   \path[line] (orig-wrapper) -- (obf-wrapper);
   \node[below=.15cm of obf-wrapper] {\sf Obfuscated Code (C + Prolog)};
    \node[block,text width=2.3cm, minimum height=4.3cm, below right=.25cm and .25cm of obf-wrapper.north west] (obf-main) {
      \begin{lstlisting}[basicstyle=\ttfamily\footnotesize]
int main() {
  ...
}

int bar() {
  ...
  x = foo(a);
  ...
  y = foo(b);
  ...
}
      \end{lstlisting}
   };

   \node[block,text width=4.7cm, below right=0cm and .5cm of obf-main.north east] (obf-foo) {
     \begin{lstlisting}[escapechar=@,basicstyle=\ttfamily\footnotesize]
int foo(int arg) {
  @\it{\textbf{\sf Declare C Local Variables;}}@
  @\it{\textbf{\sf Initialize Prolog Arguments;}}@
  @\it{\underline{\sf Start Prolog Query (\texttt{foo\_babel});}}@
  @\it{\textbf{\sf Read Prolog Return Value;}}@

  return ret;
}
      \end{lstlisting}
   };
   \node[block,text width=3.4cm, below right=0cm and .5cm of obf-foo.north east] (obf-pro) {
      \begin{lstlisting}[basicstyle=\ttfamily\footnotesize]
foo_babel(Ret, Arg)
  :- p1,
     p2,
     ...,
     pn.
      \end{lstlisting}
   };

   \node[below=.2cm of obf-foo,text width=4.8cm, text centered] (obf-foo-text) {\small \sf C-to-Prolog Wrapper};
   \node[above right=0cm and .25cm of obf-foo-text.east,text width=3.2cm, text centered] {\small \sf Prolog Predicate (translated from the original \texttt{foo})};
   \coordinate[above right=.3cm and -.2cm of obf-foo.east] (ctr);
   \path[line] (ctr) -- (obf-pro);
  \end{tikzpicture}
  \caption{The context for executing obfuscated code in \babel.}
  \label{fig:context}
\end{figure*}
}

\babel\ combines the C and Prolog runtime environments together, 
and the program starts from executing C code.
When the execution encounters an obfuscated function
(which is now a wrapper for initiating queries to the corresponding Prolog predicate),
it setups a context prior to evaluating the Prolog predicate. 
In the setup process the wrapper
allocates local variables whose addresses are referred to in the preprocessed C function. 
The wrapper then passes the addresses along with function arguments to the Prolog predicate 
through the C-to-Prolog interface provided by GNU Prolog.
\F~\ref{fig:context} illustrates how the two languages are combined.

\subsection{Customizing Prolog Engine}
Although GNU Prolog has some nice features that make it a mostly adequate candidate for
implementing \babel,
it still does not fully satisfy our requirements, thus requiring some customization. 
A notable issue about GNU Prolog is that its
interface for calling Prolog from C is not reentrant. 
This is critical because by design, users of \babel\ can freely 
choose the functions they want to obfuscate. 
To support this,
it is in general not possible to avoid stack traces that interleave C and Prolog
subroutines. We found that the non-reentrant issue results from the use of a global WAM
state across the whole GNU Prolog engine. We fixed it by maintaining 
a stack to save the WAM state before a new C-to-Prolog interface invocation and
restore the state after the call is finished.
%Due to this issue, we suspect that our current
%implementation is not threat safe and therefore cannot support multithreading.
%However, this should be not a limitation of either the Prolog programming language or \mo's methodology.

Another issue is that GNU Prolog does not implement garbage collection;
therefore memory consumption can easily explode.
This problem is not as severe as it looks because
we do not have to maintain a heap for Prolog runtime throughout the lifetime of the program.
Because we know that the life cycles of all Prolog variables are bounded by the scope of predicates, 
we can safely empty the Prolog heap when there are no pending Prolog subroutines during the execution.
Since GNU Prolog implements the heap as a large global
array and indicates heap usage with a heap-top pointer, we can empty the heap by simply 
resetting the heap-top pointer to the starting point of the heap array, which is very efficient.

\section{Evaluation}\label{sec:eval}
Collberg et al.~\cite{Collberg1998} proposed to evaluate an
obfuscation technique with respect to four dimensions: \emph{potency},
\emph{resilience}, \emph{cost}, and \emph{stealth}.
Potency measures how obscure and complex the program has become after being obfuscated.
Resilience indicates how well programs obfuscated by \babel\ can withstand reverse engineering effort,
especially automated deobfuscation.
Cost measures the execution overhead imposed by obfuscation.
Stealth measures the difficulty in detecting the existence of obfuscation, given the obfuscated binaries.
We evaluate \babel\ and observe to what extent it meets these four criteria.
%We notice that it is generally hard to evaluate obfuscation techniques.
%, especially the resilience.
%Reverse engineering could be very ad-hoc in practice and attackers can make use of any reverse engineering tool,
%some of which are even not publicly available to defenders. 
%In previous literature, obfuscation methods are usually evaluated on one or a few metrics instead of  across all four criteria. 
%We try to be comprehensive in our evaluation whenever it is feasible.

To show that our tool can effectively protect real-world software of different categories,
we apply \babel\ to six open source C programs that have been widely deployed for years or even decades.
Among the six programs, four are CPU-bound applications and the other two are IO-bound servers.
The CPU-bound applications include algebraic transformation (bzip2), integer computation (mcf), 
state machine (regexp), and floating-point computation (svm\_light).
The two IO-bound servers cover two of the most popular network protocols, i.e., FTP (oftpd) and HTTP (mongoose).
We believe that our selection is a representative evaluation set covering a wide
range of real-world software.
Table~\ref{tab:snd-eval-set} presents the details of these programs.\footnote{%
We notice that some previous work~\cite{Chen:2009:CFO:1669112.1669162} on obfuscation employed the SPEC benchmarks or GNU Coreutils,
which are also widely used in other research,
for evaluation. Unfortunately, these two software suites 
use very complicated build infrastructures. Since \babel\ needs to compile C and Prolog together,
a specialized build procedure is required. Currently our prototypical implementation of \babel\ cannot
automatically hook an existing build system, so we are not able to include SPEC or GNU Coreutils into
our evaluation.}

We define the term \textit{obfuscation level} as the percentage of 
functions obfuscated in a C program.
For example, an obfuscated bzip2 instance at the 20\% obfuscation level is a bzip2 
binary compiled from source code consisting of 80\% of the original functions in C and 
Prolog predicates translated from the other 20\% C functions by \babel.
\ignore{Formally, the number of obfuscated functions at obfuscation level $L$ is 
$\ceil[\big]{L \cdot N}$ where $N$ is total number of functions in a program.}
We achieve all obfuscation levels by randomly selecting candidates from all functions
that can be obfuscated by \babel,
but note that this random selection scheme is just for avoiding subjective picking in our research. 
In practice, \babel\ users should decide which functions
are critical and in need of protection. 
This is the same as popular commercial \vbo\ tools~\cite{code-virtualizer,VMProtect}.

\begin{table}[t]
  \centering
  \caption{Programs used for \babel\ evaluation.}\label{tab:snd-eval-set}
  \small
%\resizebox{\linewidth}{!}{
  \begin{tabular}{l|p{3.5cm}|c|c}
    \hline
    Program & Description & LoC & \# of Func. \\
    \hline
    bzip2 & Data compressor & 8,117 & 108 \\
    mcf   & Vehicle scheduler& 2,685 & 25 \\
    regexp & Regular expression engine& 1,391 & 22 \\
    svm\_light & Support vector machine& 7,101 & 103 \\
    \hline
    oftpd & Anonymous FTP server & 5,211 & 96 \\
    mongoose & Light-weight HTTP server & 5,711 & 203 \\
    \hline
  \end{tabular}
%}
\end{table}

In the evaluation, we compare \babel{} with one of the most popular commercial obfuscators, Code Virtualizer (CV)~\cite{code-virtualizer},
which is virtualization based and has been on the market since 2006.
The comparison covers all the four dimensions of evaluation, but some of the evaluation
methodologies we designed for \babel{} may not be suitable for evaluating Code
Virtualizer. For those evaluations that we consider not suitable for CV, we will explain the reasons
and readers should be cautious in interpreting the data.

\begin{table*}[t]
\centering
\caption{Program complexity before and after \babel\ obfuscation at 30\% obfuscation level} \label{table:potency-binary}
\setlength\tabcolsep{1pt}
 \begin{tabular}{l|P{4em}P{4em}P{2.3em}|P{4em}P{4em}P{2.3em}|P{4em}P{4em}P{2.3em}|P{4em}P{4em}P{2.3em}|P{4em}P{4em}P{2.3em}}
   \hline
   \multirow{2}{*}{Program} & \multicolumn{3}{c|} {\# of Call Graph Edges} & \multicolumn{3}{c|} {\# of CFG Edges} & \multicolumn{3}{c|} {\# of Basic Blocks} & \multicolumn{3}{c|} {Cyclomatic Number} & \multicolumn{3} { c } {Knot Count} \\ \cline{2-16}
                & Original & \babel & Ratio & Original & \babel & Ratio & Original &  \babel & Ratio & Original & \babel & Ratio & Original & \babel & Ratio \\
   \hline
   bzip2& 353 & 5964 & 16.9 & 5382 & 19771 & 3.7 & 3528 & 17078 & 4.8 & 1856 & 2695 & 1.5 & 3120 & 12396 & 4.0 \\
   mcf& 78 & 5449 & 69.9 & 854 & 14233 & 16.7 & 583 & 13086 & 22.4 & 273 & 1149 & 4.2 & 153 & 8792 & 57.5 \\
   regexp& 72 & 5276 & 73.3 & 855 & 13290 & 15.5 & 591 & 11802 & 20.0 & 266 & 1490 & 5.6 & 1135 & 9530 & 8.4 \\
   svm& 511 & 6739 & 13.2 & 5375 & 20752 & 3.9 & 3545 & 18533 & 5.2 & 1832 & 2221 & 1.2 & 2972 & 11521 & 3.9 \\
   \hline
   oftpd& 455 & 5810 & 12.8 & 2035 & 15501 & 7.6 & 1667 & 14422 & 8.7 & 370 & 1081 & 2.9 & 1277 & 9911 & 7.8 \\
   mongoose& 1027 & 6762 & 6.6 & 2788 & 17115 & 6.1 & 2086 & 16079 & 7.7 & 704 & 1038 & 1.5 & 493 & 9491 & 19.3 \\
   \hline
   \hline
   Geom.Mean.& 279.2 & 5972.7 & 21.4 & 2220.4 & 16555.5 & 7.5 & 1570.2 & 14987.8 & 9.5 & 633.0 & 1502.4 & 2.4 & 1002.3 & 10199.1 & 10.2 \\
   \hline
 \end{tabular}
\end{table*}
\begin{table*}[t]
\centering
\caption{Program complexity before and after Code Virtualizer (CV) obfuscation at 30\% obfuscation level} \label{table:potency-cv}
\setlength\tabcolsep{1pt}
 \begin{tabular}{l|P{4em}P{4em}P{2.3em}|P{4em}P{4em}P{2.3em}|P{4em}P{4em}P{2.3em}|P{4em}P{4em}P{2.3em}|P{4em}P{4em}P{2.3em}}
   \hline
   \multirow{2}{*}{Program} & \multicolumn{3}{ c| } {\# of Call Graph Edges} & \multicolumn{3}{ c| } {\# of CFG Edges} & \multicolumn{3}{c|} {\# of Basic Blocks} & \multicolumn{3}{ c| } {Cyclomatic Number} & \multicolumn{3} { c } {Knot Count} \\ \cline{2-16}
                & Original & CV & Ratio & Original & CV & Ratio & Original &  CV & Ratio & Original & CV & Ratio & Original & CV & Ratio \\
   \hline
   bzip2& 353 & 261 & 0.7 & 5382 & 3868 & 0.7 & 3528 & 2826 & 0.8 & 1856 & 1044 & 0.6 & 3120 & 713 & 0.2 \\
   mcf& 78 & 34 & 0.4 & 854 & 461 & 0.5 & 583 & 329 & 0.6 & 273 & 134 & 0.5 & 153 & 68 & 0.4 \\
   regexp& 72 & 66 & 0.9 & 855 & 525 & 0.6 & 591 & 377 & 0.6 & 266 & 150 & 0.6 & 1135 & 603 & 0.5 \\
   svm& 511 & 357 & 0.7 & 5375 & 3267 & 0.6 & 3545 & 2358 & 0.7 & 1832 & 911 & 0.5 & 2972 & 302 & 0.1 \\
   \hline
   oftpd& 455 & 390 & 0.9 & 2035 & 1727 & 0.8 & 1667 & 1435 & 0.9 & 370 & 294 & 0.8 & 1277 & 542 & 0.4 \\
   mongoose& 1027 & 585 & 0.6 & 2788 & 2063 & 0.7 & 2086 & 1638 & 0.8 & 704 & 427 & 0.6 & 493 & 442 & 0.9 \\
   \hline
   \hline
   Geom.Mean.& 279.2 & 190.4 & 0.7 & 2220.4 & 1489.0 & 0.6 & 1570.2 & 1117.0 & 0.7 & 633.0 & 365.9 & 0.6 & 1002.3 & 358.3 & 0.3 \\
   \hline
 \end{tabular}
\end{table*}

\subsection{Potency}\label{sec:potency}
We use two groups of static metrics to show how much complexity \babel\ has
injected into the obfuscated programs.
The first group consists of basic statistics about the call graph and control-flow graph (CFG),
including the number of edges in both graphs and the number of basic blocks.
These metrics have been used to evaluate obfuscation techniques
in related work~\cite{Chen:2009:CFO:1669112.1669162}.

In addition to basic statistics,
we also calculate two metrics used to quantify program complexity,
proposed by historical software engineering research.
The measures are the cyclomatic number~\cite{mccabe_complexity_1976} and the knot count~\cite{woodward_measure_1979}.
Both metrics reflect Gilb's statement that logic complexity is a measure of the degree of decision
making within a system~\cite{gilb1977software}. They also have been considered for
evaluating obfuscation effects~\cite{Collberg1998}.
The cyclomatic number is defined as $e-n+2$ where $e$ and $n$ are the numbers of edges and vertices 
in the CFG. Intuitively, the cyclomatic number represents the amount of
decision points in a program~\cite{Conte:1986:SEM:6176}.
The knot count is the amount of edge crossings in the CFG
when all nodes are placed linearly and all edges are drawn on the same side.

Table~\ref{table:potency-binary} shows the comparison between binaries with and without
\babel\ obfuscation on the complexity measures we have chosen, at the obfuscation level of 30\%.%
\shortlong{%
\footnote{We choose 30\% because it is a more realistic configuration in practice.
In case that readers are interested in \babel's potency at other other obfuscation levels,
please refer to an extended version of this paper~\cite{longversion}.}}%
{We choose 30\% because it is a more realistic configuration in practice.
Readers can refer to the appendix for 
potency evaluation data at other obfuscation levels, i.e., from 10\% to 50\%.
}
To be conservative, by the time of measurement we have stripped the code belonging to the GNU Prolog runtime itself,
so the extra complexity (if there is any) should be purely credited to \babel's obfuscation.
We use IDA Pro~\cite{ida}, an advanced commercial reverse engineering tool,
to disassemble the binary and generate call graphs and control-flow graphs.
As can be seen, the obfuscated binaries have a significant advantage on all metrics.
The geometric mean of all six programs shows that \babel\ can vastly increase program complexity.
Note that different from static complexity produced by obfuscation methods using opaque predicates,
the additional control-flow branches injected by \babel\ are ``real'' in the sense that all branches can be feasible
at run time.

Our potency evaluation is not an ideal methodology for measuring the same aspect of Code Virtualizer. 
The reason is that Code Virtualizer transforms binary instructions into bytecode and
the reverse engineering tool we use, i.e., IDA Pro, is incapable of handling this
situation.
Table~\ref{table:potency-cv} shows the complexity of binaries with and without
Code Virtualizer obfuscation, obtained in the same way as Table~\ref{table:potency-binary}.
The data suggests that after Code Virtualizer is applied, the complexity of the protected binaries has a 
notable decrease compared to the original ones. In reality, this is a consequence of the aforementioned issue.
Because IDA Pro is unable to analyze the re-encoded functions, the potency evaluation inevitably misses a significant
portion of the complexity.

\subsection{Resilience}\label{sec:resilience}
%The resilience of an obfuscation technique measures how much effort is
%required for \textit{automated tools} to effectively reduce its potency.
In general, the resilience of an obfuscation technique is hard to assess
because reverse engineering can be a very ad-hoc process.
On the other hand, very few practical deobfuscation tools are publicly
available to the community. Because of these difficulties, some previous work
on software obfuscation failed to evaluate resilience properly.
In our evaluation, we choose binary diffing to assess
\babel's resilience after intensive investigation, although it does not mean binary diffing is the only
deobfuscation technique.

Binary diffing is a commonly used reverse engineering technique
which calculates the similarity between two binaries.
We consider binary diffing a deobfuscation technique
because it reveals the connection from an obfuscated program to its original version.
Given a program binary and its obfuscated version,
if a binary diffing tool reports high similarity score for the comparison (ignoring potential false positives),
then in some sense the differ has successfully undone the obfuscation effect.
Most historical work on deobfuscation known to us
uses similarity-based metrics to evaluate the effectiveness of their techniques~\cite{Sharif2009,Coogan2011,yadegari2014generic}.
\longversion{
Deciding the similarity of untrusted programs, especially binaries, has been such an important topic in computer security
that DARPA has initiated the four year, \$43 million Cyber Genome Program to support related
research~\cite{cgp}.
}

Binary similarity can be calculated based on either syntax or semantics.
The syntax mostly refers to the control flows of the binary and
syntax-based binary diffing usually takes a graph-theoretic approach which
compares the call graphs of two binaries and further the control flow graphs of pairs of functions between two binaries, looking for
any graph or subgraph isomorphism. The intuition is,
if two binaries have similar call graphs, the functions located at corresponding nodes in the call graph isomorphism
are likely to be similar ones;
if two functions have similar control flows, they are likely to implement the same
computation logic.

On the other hand, semantics-based binary diffing focuses more on the observable behavior of the binaries.
There are various ways to describe program behavior, e.g., the post- or pre-condition of a
given chunk of code and certain effects the code commits such as system calls.
If two binaries have matched behavior, a semantics-based binary diffing tool will consider
them similar.

In general, syntax-based similarity is less strict than semantics-based similarity.
Relatively,
syntax-based differs tend to report more false positives while semantics-based differs
tend to get more false negatives. To avoid bias as much as possible in the evaluation,
we pick binary differs of both kinds to test \babel's resilience to reverse engineering.
We employ CoP~\cite{LUO} and BinDiff~\cite{bindiff}, of which
CoP is a semantics-based binary differ and BinDiff is syntax based~\cite{\shortlong{thomas2005graph}{flake04,thomas2005graph}}.
To measure \babel's resilience to a differ,
we randomly pick 50\% functions from each program in Table~\ref{tab:snd-eval-set}, obfuscate them with \babel,
and then launch the differs to calculate the similarity between the original and obfuscated functions.

\subsubsection{Resilience to Semantics-Based Binary Diffing}
CoP, a ``\textit{semantics-based obfuscation-resilient}''
binary similarity detector~\cite{LUO}, is currently one of the state-of-the-art semantics-based binary diffing tools.
The detection algorithm of CoP is founded on the concept of ``longest common
subsequence of semantically equivalent basic blocks.''
By constructing symbolic formulas to describe the input-output relations of basic blocks,
CoP checks the semantic equivalence of two basic blocks with a theorem prover.
It is reported that this new binary diffing technique can
defeat many traditional obfuscation methods.
CoP is built upon several cutting-edge techniques in the field of reverse engineering,
including the binary analysis toolkit BAP~\cite{Brumley:2011:BBA:2032305.2032342} and
the constraint solver STP~\cite{Ganesh:2007:DPB:1770351.1770421}.
CoP defines the similarity score as the number of matched basic blocks divided by
the count of all basic blocks in the original function.

\begin{figure}[t]
\includegraphics[width=\linewidth,trim=0 3em 0 2.5em]{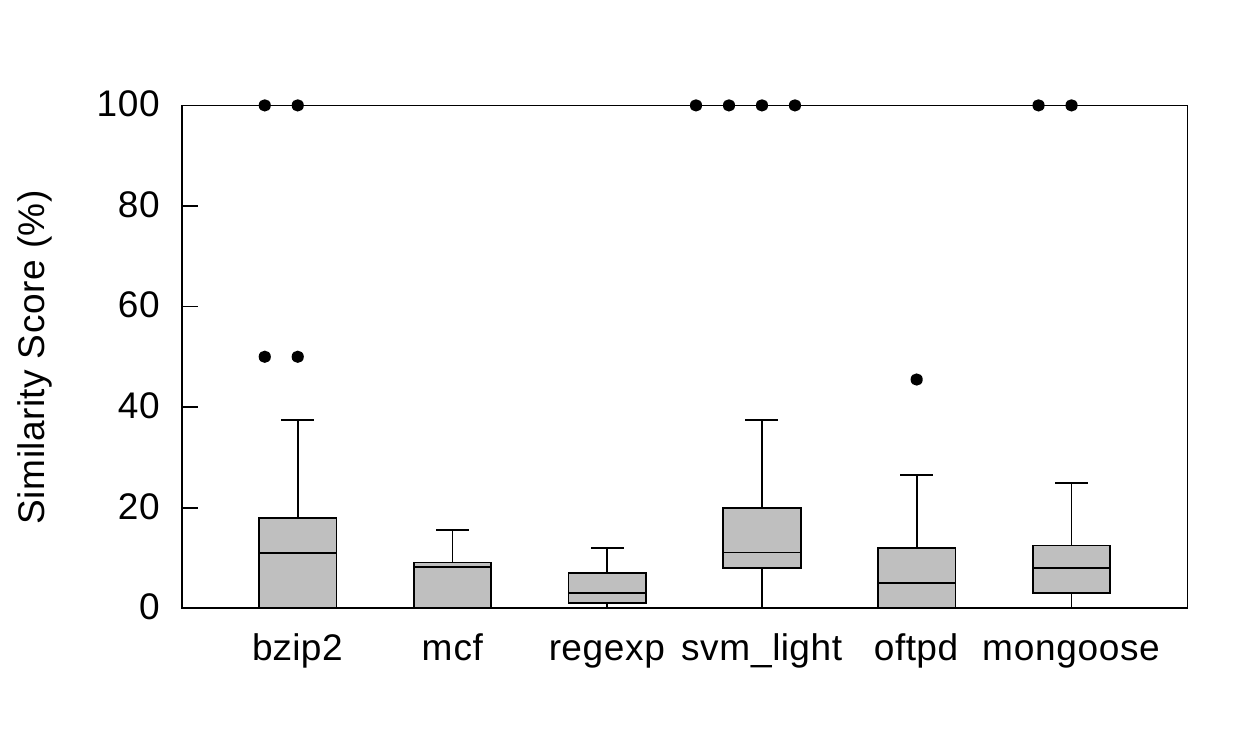}
\caption{Distributions of similarity scores between the original and \mbox{\babel}-obfuscated functions in the evaluated programs.}
\label{fig:cop}
\end{figure}

\F~\ref{fig:cop} is the box plot showing the distribution of similarity scores.
For all programs, the third quartile of the scores is below 20\%.
Considering that the original paper of CoP reports over 70\% similarity in most
of their tests on transformed or obfuscated programs,
the scores calculated from \babel-obfuscated functions
are not convincing evidence of similarity.
One may notice that there are a few outliers in \F~\ref{fig:cop}, i.e., the similarity scores for some functions can reach 100\%.
These functions are all ``simple'' ones, namely they
have only one basic block and very few lines of C code.
With the presence of false positives, it is very likely that the binary differ can report
100\% similarity for these functions, according to CoP's similarity score definition.

\subsubsection{Resilience to Syntax-Based Binary Diffing}
BinDiff is a proprietary
syntax-based binary diffing tool which is the de facto industrial standard with wide availability.
It has motivated the creation of several academia-developed binary differs such as
BinHunt~\cite{gao_binhunt:_2008} and its successor iBinHunt~\cite{iBinhunt}.

Given two binaries, BinDiff will give a list of function pairs that are considered similar based on
a set of different algorithms. %BinDiff have a pool of algorithms used to calculate similarity.
In addition to the similarity level like CoP reports, BinDiff also reports its ``confidence'' on the results,
based on which algorithm is used to get that score. It is not completely clear to us
how each of BinDiff's algorithms works and how BinDiff ranks the confidence level.
Therefore, we report how many obfuscated functions are correctly matched to their
originals by BinDiff regardless of the similarity score and the confidence. This makes sure
\babel\ does not take any unfair advantage over its opponent in the evaluation of performance.
In other words, the results reported here indicate a lower bound of \babel's resilience to syntax-based
binary diffing.

\begin{table}[t]
  \centering
  \caption{Function matching result from BinDiff on \babel-obfuscated programs}
  \begin{tabular}{lccc}
    \hline
    Program & \# of Obfuscated & \# of Matched & Match Rate \\
    \hline
    bzip2 & 54 & 6 & 11.11\% \\
    mcf & 13 & 7 & 53.85\% \\
    regexp & 11 & 3 & 27.27\% \\
    svm\_light & 52 & 10 & 19.23\% \\
    \hline
    oftpd & 48 & 4 & 8.33\% \\
    mongoose & 102 & 39 & 38.24\% \\
    \hline
    \hline
    Overall & 280 & 69& 24.64\% \\
    \hline
  \end{tabular}
  \label{tab:bindiff}
\end{table}

\begin{figure}[t]
\includegraphics[width=\linewidth,trim=0 3em 0 2.5em]{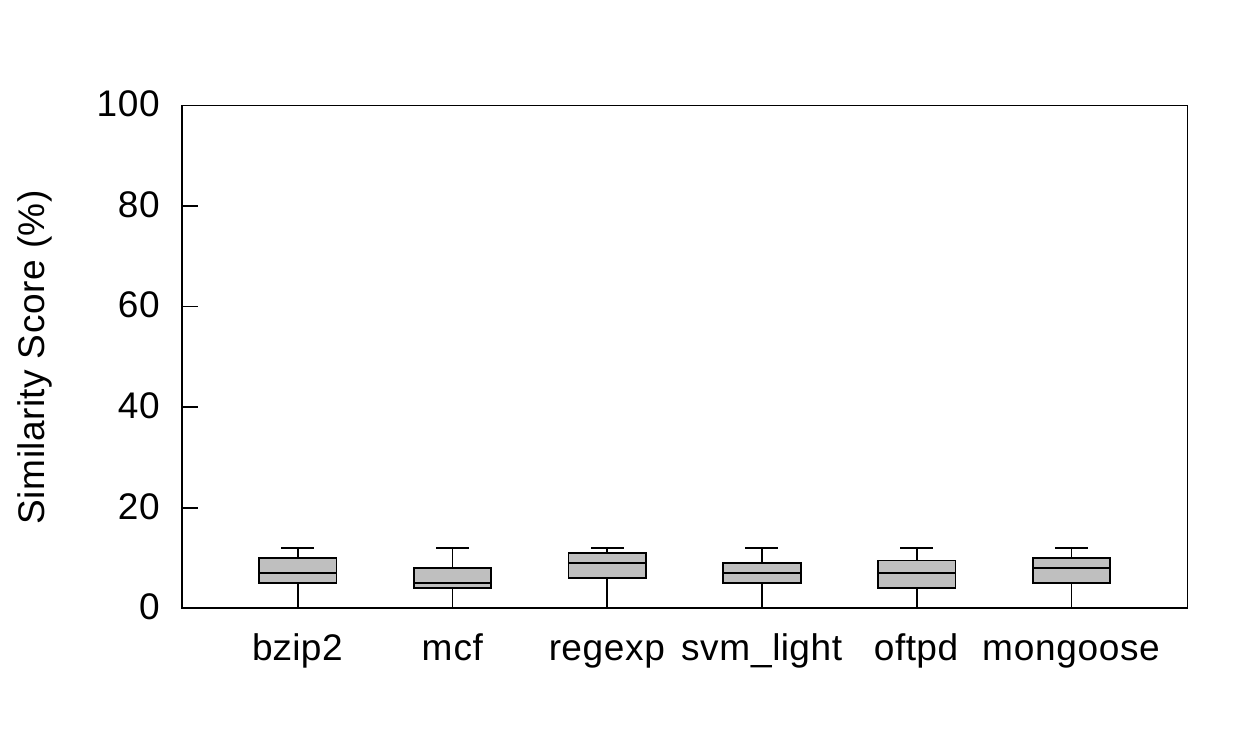}
\caption{Distributions of similarity scores between the original and CV-obfuscated functions in the evaluated programs.}
\label{fig:cop-cv}
\end{figure}
\begin{table}[t]
  \centering
  \caption{Function matching result from BinDiff on CV-obfuscated programs}
  \begin{tabular}{lccc}
    \hline
    Program & \# of Obfuscated & \# of Matched & Match Rate \\
    \hline
    bzip2 & 54 & 7 & 12.96\% \\
    mcf & 13 & 4 & 30.77\% \\
    regexp & 11 & 0 & 0.00\% \\
    svm\_light & 52 & 1 & 1.92\% \\
    \hline
    oftpd & 48 & 2 & 4.17\% \\
    mongoose & 102 & 11 & 10.78\% \\
    \hline
    \hline
    Overall & 280 & 25& 8.93\% \\
    \hline
  \end{tabular}
  \label{tab:bindiff-cv}
\end{table}

Table~\ref{tab:bindiff} shows how many obfuscated functions in each program are matched
(although some of them get low similarity scores or confidence).
Since BinDiff can match functions solely based on their coordinates in the call graphs,
two functions can be matched even if they have totally different semantics. This explains
why BinDiff can achieve a relatively high matching rate for mongoose, because mongoose has
the largest number of functions and potentially has a more iconic call graph.
Nevertheless, only 26.22\% of the obfuscated functions are matched by BinDiff over all six programs.
Note that matching does not yet imply successful deobfuscation or recovery of
program logic, especially for syntax-based binary differs.
In that sense, we believe \babel's performance is satisfying.

\subsubsection{Comparing \babel\ with Code Virtualizer}
We present
Code Virtualizer's resilience to CoP and BinDiff in \F~\ref{fig:cop-cv} and
Table~\ref{tab:bindiff-cv}, respectively. The data are obtained in experiments of which the settings are
consistent with the resilience evaluation on \babel.
Based on the data, it seems that Code Virtualizer is more resilient to CoP and BinDiff than \babel.
However, as aforementioned in the potency evaluation,
reverse engineering binaries protected by virtualization-based obfuscators like
Code Virtualizer requires
specialized approaches. Since neither CoP nor BinDiff is made aware of the fact that
the obfuscated parts of the binaries have been transformed from code to data,
their poor performance is not a surprising result.
After all, a major weakness of virtualization-based obfuscation is that
although the original program may be well obfuscated, the virtual machine itself is
still exposed to attacks. By reverse engineering the logic of the virtual machine
and revealing the encoding format of the bytecode, attackers can effectively deobfuscate
the protected binaries~\cite{Sharif2009,yadegari2014generic}.

\subsection{Cost}\label{sec:cost}
\begin{comment}
\longversion{
We measure the cost of \babel\ obfuscation by its impact on the size and speed of executables.
\subsubsection{Binary Size}\label{sec:size}

\begin{figure}[t]
\centering
\includegraphics[width=\linewidth,trim=0 1.5em 0 0]{plot/size.pdf}
\caption{Binary size expansion of obfuscated program. 0\% obfuscation level means only the Prolog runtime is integrated without actually translating any original C code.} \label{fig:cost-size}
\end{figure}

For all evaluated programs, we calculate binary sizes at different obfuscation levels.
The growing trend is displayed in \F~\ref{fig:cost-size}.
Roughly, \babel\ expands binary sizes linearly with respect to obfuscation level and original binary size.
The x axis does not start from 0\% because we want to present the cost of just
embedding the obfuscation environment. \ignore{(the GNU Prolog runtime system) without actually obfuscating any code.}
The data show that the space occupied by the runtime system itself is about constant.
\ignore{Overall, the extra space needed to store programs protected by \babel\ is modest and acceptable given 
the storage capacity of today's computing environments.}

\subsubsection{Execution Speed}
}
\end{comment}

\begin{table*}[t]
\setlength\tabcolsep{3pt}
  \caption{Time overhead introduced by \babel\ and Code Virtualizer (CV)}
  \label{tab:cost-time}
\centering
{
  \begin{tabular}{l|c|cc|c|cc|c|cc|c|cc|c|cc}
    \hline
    \multirow{3}{*}{Program}& \multicolumn{3}{c|}{10\% obfuscated} & \multicolumn{3}{c|}{20\% obfuscated} & \multicolumn{3}{c|}{30\% obfuscated} & \multicolumn{3}{c|}{40\% obfuscated} & \multicolumn{3}{c}{50\% obfuscated} \\ 
    \cline{2-16}
    &\multirow{2}{*}{Coverage} & \multicolumn{2}{c|}{Slowdown }& \multirow{2}{*}{Coverage} &\multicolumn{2}{c|}{Slowdown }& \multirow{2}{*}{Coverage} &\multicolumn{2}{c|}{Slowdown }&\multirow{2}{*}{Coverage} &\multicolumn{2}{c|}{Slowdown }&\multirow{2}{*}{Coverage} &\multicolumn{2}{c}{Slowdown}\\
    \cline{3-4}     \cline{6-7}      \cline{9-10} \cline{12-13} \cline{15-16}
                           & & \babel & CV & & \babel &  CV & & \babel & CV &&\babel&CV&&\babel&CV\\
    \hline
    bzip2                  & 0.00\% & 1.5   &1.9  & 0.00\% &1.5   &1.9 & 27.78\% &8.0   & $\times$ &33.34\%&100.9&$\times$&33.34\%&105.8&$\times$\\
    mcf                    & 0.66\%& 1.0   & 12.0 & 4.30\% &6.9   & 52.4 & 15.54\% & 65.8  &$\times$&69.33\%&135.9&$\times$&83.33\%&169.3&$\times$\\
    regexp                 & 18.33\% & 138.5  &660.9& 19.74\% &173.8  &834.8& 19.74\%&198.7  &1122.2&28.91\%&285.9&$\times$&28.91\%&288.8&$\times$\\
    svm\_light             & 13.33\% & 1.0   & 58.0 & 20.00\% &4.0   & $\times$&26.67\% &4.1   &$\times$   &26.67\%&4.1&$\times$&33.33\%&11.8& $\times$\\
    \hline
    oftpd                  & - & 1.0   &1.0  & - &1.1   &1.1  &- & 1.1  & 1.1   &-&1.1&$\times$&-&1.1&$\times$\\
    mongoose               & - & 1.0   & 1.4 & - &1.2   &8.8  &- &1.6   & 9.3   &-&1.7&$\times$&-&2.1&$\times$\\
    \hline
  \end{tabular}
}
  \vspace{.05in}\\
  {\scriptsize  Coverage denotes the percentage of CPU time taken by the obfuscated functions in the execution of the original programs (not available for IO-bound servers). The percentage provided is a lower bound because some functions may be inlined into others. $\times$ indicates that the corresponding test failed due to program crash or incorrect output.}
\end{table*}
We measure execution slowdown introduced by \babel\ from the obfuscation level of 10\% to 50\%.
We use the test cases shipped with the obfuscated software as the performance test input for CPU-bound applications.
For FTP server oftpd, we transfer 10 files ranging from 1KB to 128MB.
For HTTP server mongoose, we sequentially send 100 quests for a 2.5KB HTML page. 
We conduct the experiments on a desktop with Xeon E5-1607 3.00GHz CPU and 4GB memory running 64-bit Ubuntu 12.04 LTS, 
over a 1Gbps Ethernet link. 
We run each test 10 times and report the average slowdown. 
For servers, time spent on network communication is included by our measurement.

Unlike potency and resilience, we can easily conduct a fair comparison between \babel\ and Code Virtualizer
on execution overhead.
In the comparison, we configure Code Virtualizer to minimize obfuscation strength and maximize execution speed.
The implication of our comparison setting is that, 
if \babel\ can achieve comparable or better performance than a mature commercial product, 
the runtime overhead introduced by \babel\ should be acceptable for practical use. 
To show that the functions we obfuscate are non-trivial, we use \texttt{gprof} to get their performance coverage, i.e.,
the percentage of CPU time taken by the obfuscated functions in the execution of the original programs. Note that the percentage
we obtain is just a lower bound because some functions may be inlined and \texttt{gprof} will contribute their execution time
to other functions. We are only able to get the coverage data for the four CPU-bound applications, because the CPU time spent by the two original
server programs is too short for meaningful profiling. Profiling server programs usually requires dedicated profilers and we are unaware of
the existence of such tools for the two server programs we picked.

Table~\ref{tab:cost-time} gives the experiment results, showing that \babel\ outperforms Code Virtualizer in most of the cases we tested.
In particular, \babel's obfuscation is more reliable in the sense that the obfuscated programs exit normally and give correct output on test input, 
while Code Virtualizer fails to provide reliable obfuscation on most of the tested programs when obfuscation level reaches 40\%.
Both \babel\ and Code Virtualizer impose considerably high performance overhead after the obfuscation level reaches 30\%, for many of the CPU-bound
applications. This is expected, because in general such heavy-weight obfuscation methods should be avoided when protecting program hot spots~\cite{cv-manual}.
In the evaluation, the coverage of many applications exceeds 30\% after the obfuscation level reaches 50\%, which is rarely the case if \babel\ and Code Virtualizer
are to be deployed in practice. 
Regardless, the key point of this evaluation is to demonstrate that \babel's performance cost is lower than an industry-quality obfuscator which shares certain
similarity with \babel{}.

\subsection{Stealth}\label{sec:stealth}
\newcommand{\distexample}{svm\_light}

By evaluating stealth we investigate whether \babel\
introduces abnormal statistical characteristics to the obfuscated code.
In the stealth evaluation, we pick the 30\% obfuscation level.
\begin{figure}[t]
\includegraphics[width=1\linewidth,trim=2.5em 2.5em 2.5em 0em]{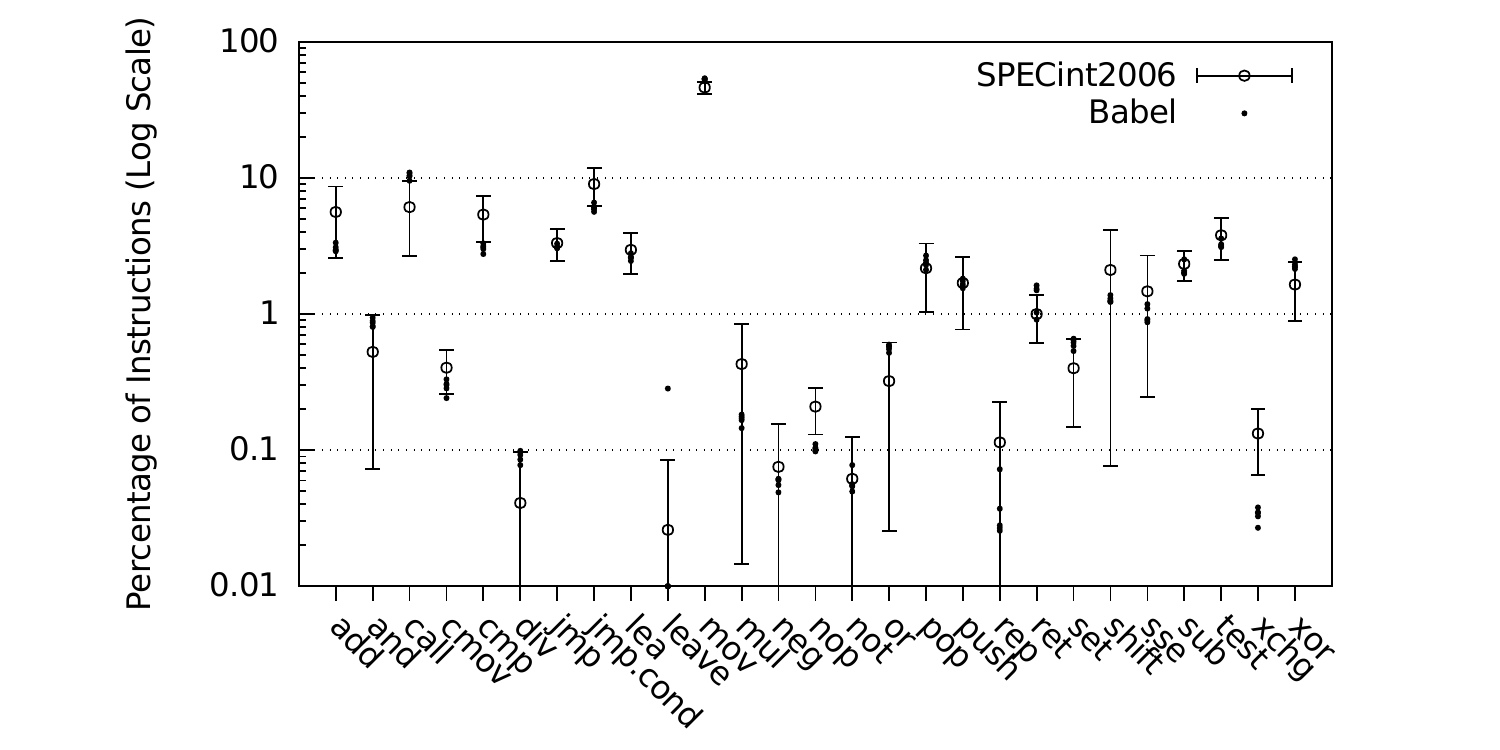}
\caption{Instruction distributions of SPECint2006 programs (mean and standard deviation) and \babel-obfuscated integer programs.}
\label{fig:inst-dist}
\vspace{20pt}
  \centering
  \includegraphics[width=1\linewidth,trim=2.5em 2.5em 2.5em 0em]{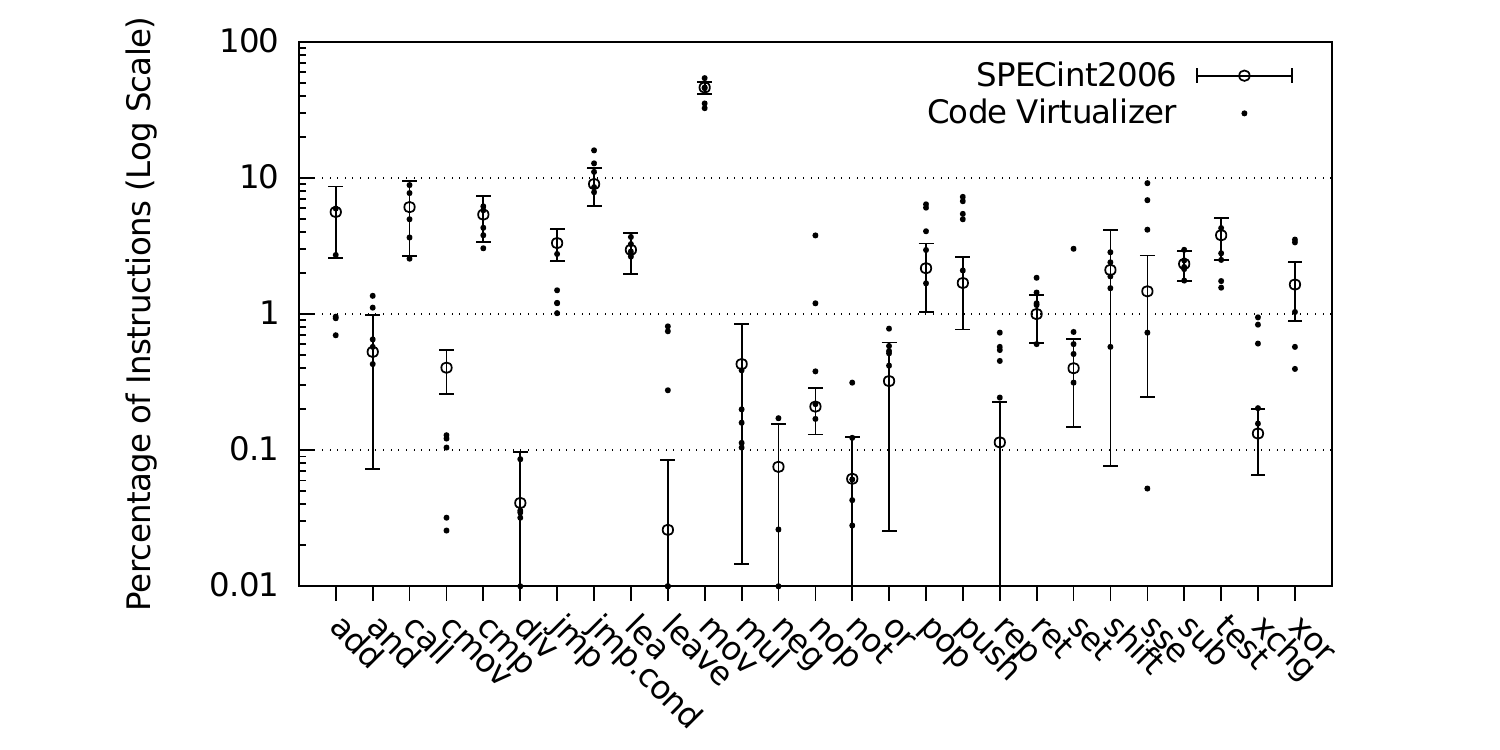}
  \caption{Instruction distributions of SPECint2006 programs (mean and standard deviation) and CV-obfuscated programs}
  \label{fig:stealth-cv}
\end{figure}

Some previous work measures obfuscation stealth by the byte entropy of program binaries~\cite{Wu:2010:MNA:1866307.1866368},
for byte entropy has been used to detect packed and encrypted binaries~\cite{entropy07}.
\longversion{
%% \begin{comment}
%% \F~\ref{fig:byte-entropy} displays the byte entropy for the original, \babel{}-protected, and CV-protected versions of
%% each evaluated program. As can be seen, the CV-protected binaries have a notable statistical anomaly while the byte
%% entropy of the \babel{}-protected versions are very close to those of the original binaries.
%% \begin{figure}
%% %\includegraphics[width=1\linewidth,trim=2.5em 2.5em 2.5em 0em]{}
%% \caption{Byte entropy comparison among the original, \babel{}-protected, and CV-protected binaries.}
%% \label{fig:byte-entropy}
%% \end{figure}
%% \end{comment}
}
Since \babel{} does not re-encode original binary code, possessing normal byte entropy 
may not be a strong evidence of stealth. Therefore,
we employ another statistical feature, the distribution of instructions, to evaluate \babel{}.
This metric has also been employed by previous work~\cite{Popov2007,Chen:2009:CFO:1669112.1669162}.
To tell whether \babel{}-obfuscated programs have abnormal instruction distributions,
we need to compare them with normal programs. Since the scale of programs used in our evaluation is relatively
small, we select the SPEC2006 benchmarks as
the representatives of normal programs. We group common x86 instructions into 27 classes and
calculate the means and standard deviations of percentages for each group within SPEC2006 programs.
Since integer programs and floating-point programs have different distributions,
we only compare bzip2, mcf, regexp, oftpd, and mongoose with SPECint2006.
%then we compare svm\_light with SPECfp2006.

\F~\ref{fig:inst-dist} presents the comparisons for integer programs.
For the majority of instruction groups, their distributions in \babel-obfuscated programs
fall into the interval of normal means minus/plus normal standard deviations. There are some exceptions such as \texttt{mov}, \texttt{call}, \texttt{ret}, \texttt{cmp}, and \texttt{xchg}. 
However, their distributions are still bounded by the minimum and maximum of SPEC distributions (not shown in the figure).
Hence, we believe these exceptions are not significant enough to conclude that \babel-obfuscated programs are abnormal in terms of instruction distribution.
%For the floating-point program comparison, the observation is the same. Due to limited space, we do not present the detailed data.

Meanwhile for binaries obfuscated by Code Virtualizer, the instruction distributions are significantly more deviant, as shown by \F~\ref{fig:stealth-cv}. It should be noted that when we tried to disassemble binaries processed by Code Virtualizer, the disassembler reports hundreds of decoding errors,
presumably because Code Virtualizer transforms legal instructions to bytecode which cannot be correctly decoded by the disassembler.
Nevertheless, this disassembly anomaly itself can also be strong evidence of obfuscation. Overall, the experiments indicate that \babel{} has better stealth performance than Code Virtualizer.

There may be of a concern that solely the existence of a Prolog execution environment 
can be the evidence of obfuscation.
This can be tackled by developing a customized Prolog engine.
Previous work has shown that a Prolog engine can be implemented with less
than 1,000 lines of Pascal or C code~\cite{Kamin:1990:PLI:78092,flit}.

\section{Discussion}\label{sec:discussion}
\subsection{Generalizing \MO{}}
Although it is usually quite challenging to translate programs in one language to another
language with very different syntax, semantics, and execution models,
many of the obstacles can be circumvented when the translation is for obfuscation
purposes and not required to be complete. In our translation from C to Prolog,
we designate the task of supporting C memory model, which is one of the most challenging
issues in translating C, partially to the C execution environment itself. This solution is
not feasible in general-purpose language translations. 
Meanwhile, some of our translation techniques are universally applicable to a class of
target languages that share certain similarities. For example, the control flow regularization
methods we proposed can be adopted when translating C to many declarative programming languages. 
We believe that \mo\ has the potential to be made a general framework that supports
various source and target languages.

\subsection{Multithreading Support}
Our current implementation of \babel{} does not support C multithreading,
and the main reason is that some components of GNU Prolog are not thread safe.
Since GNU Prolog is a Prolog implementation for research and educational use,
some language features are not supported. However, many other Prolog
implementations that are more mature can indeed support multithreading well~\cite{multithread}. 
By investing enough engineering effort, we should be able to improve the implementation
of GNU Prolog and ensure that it supports concurrent programming.
Therefore, we do not view the current limitation as a fundamental one.

\subsection{Randomness}
Some obfuscation techniques improve the security strength by introducing randomness.
For example, the virtualization-based obfuscators usually randomize the encoding of their
virtual instruction set~\cite{eilam2011reversing} so that attackers cannot crack all randomized binaries by 
learning the encoding of a single instance. Although this randomization is ineffective
once attackers learned how to systematically crack the virtual machine itself,
the idea of randomization does have some value.

Our current design of \mo{} does not explicitly feature any randomness. However,
since \mo{} is orthogonal to existing obfuscation techniques, it can be stacked with those
techniques that do introduce randomness. \Mo{} itself has the potential to feature
randomness as well. One promising direction could be making some of the foreign language
compilation strategies undeterministic. Previous
research~\cite{egele_blanket_2014} has shown that mutating compilation configurations can
effectively disrupt some deobfuscation tools.

\subsection{Defeating \MO}
%It is natural to raise the question that how difficult it is to
%reverse engineer the Prolog language in general, since this directly relates to the
%security foundation of \mo. 
%As we have previously emphasized
In general, \mo{} is open design and does not rely on any secrets,
although it can be combined with other secret-based obfuscation methods. 
All of our justification on the security strength of \mo{} assumes
that attackers do possess the knowledge that we have translated C into Prolog.
Indeed, with this knowledge attackers can choose to convert the binary to Prolog
first rather than directly getting back to C. Either way,
attackers will face severe challenges.

We would like to emphasize again
that we do not argue it is impossible to defeat \mo{}.
Instead, we argue that Prolog is more difficult to crack than C, in the \mo\ context.
As long as a \mbox{\babel}-translated Prolog predicate is compiled as native code,
recovering it to a high-level program representation faces all the difficulties encountered in C reverse
engineering, including the hardness of disassembly and analysis~\cite{uroboros}.
In \S\ref{sec:background} we revealed the 
deep semantics gap between Prolog source code and its low-level implementation.
Thanks to this gap, we expect that recovering the computation logic of native code compiled
from Prolog-translated C source code will consume a significant amount of reverse engineering
effort.

What makes defeating \mo\ even more challenging is that, the obfuscated code is not only a
plain combination of normal Prolog plus normal C but a tangled mixture of both. The execution of
obfuscated programs will switch back and forth between the two language environments and there
will be frequent interleaving of different memory models (see \S\ref{sec:design}).
This also imposes challenges to reverse engineering.

There is another point that grants \mo\ the potential to significantly delay reverse
engineering attacks.
As stated in \S\ref{sec:benefits}, \mo\ is not limited to Prolog. There are 
many other programming languages that we can misuse for protection. By mixing these languages
in a single obfuscation procedure, the difficulty of reverse engineering will be further increased.

\section{Related Work}\label{sec:related}
\subsection{Programming Language Translation}
People seek to translate one programming language to another, especially from source to source, 
for portability, re-engineering, and security purposes.
The source-to-source translation
from C/C++ to Java is one of the most extensively explored topics in this field, 
leading to tools such as C2J~\cite{laffra2001c2j}, C++2Java~\cite{c++2java}, and Cappuccino~\cite{Buddrus:1998:CMC:330560.331015}, etc.
Trudel et al.~\cite{trudel_c_2012} developed a converter that translates C to Eiffel, another object-oriented programming language.
A tool called Emscripten can translate LLVM intermediate representation to JavaScript~\cite{zakai_emscripten:_2011}.
Since C/C++/Objective-C source code can be compiled into LLVM intermediate representation, 
Emscripten can also be used as a source-to-source translator without much additional effort.
The C-to-Prolog translation introduced in this paper is partial since we need to keep the
original C execution environment; however, our translation is for software obfuscation and
being partial is not a limitation. Instead, we show that for our purpose, 
many technical issues commonly seen in programming language translation can be either
addressed or circumvented.

\subsection{Obfuscation and Deobfuscation}
Software obfuscation can be on either source level or binary level.
For source code obfuscation, 
Sharif et al.~\cite{Sharif2008}
encrypted equality conditions that depend on input data with some
one-way hash functions. The evaluation shows that it is virtually
impossible to reason about the inputs that satisfy the equality
condition with symbolic execution. Moser et al.~\cite{Moser2007}
demonstrate that opaque predicates can effectively hide control
transfer destination and data locations from advanced malware
detection techniques. 

Obfuscation-oriented program transformations can also be performed at the binary level.
Popov et al.~\cite{Popov2007} obfuscate programs by replacing control
transfers with exceptions, implementing real control transfers in
exception handling code, and inserting redundant junk transfers after the exceptions. 
Mimimorphism~\cite{Wu:2010:MNA:1866307.1866368} transforms a malicious binary
into a mimicry benign program, with statistical and semantic
characteristics highly similar to the mimicry target. As a
result, obfuscated malware can successfully evade statistical anomaly
detection. Chen et al.~\cite{Chen:2009:CFO:1669112.1669162} propose
a control-flow obfuscation method making use of Itanium processors' architectural support for
information flow tracking. In detail, they utilize the deferred exception tokens in
Itanium processor registers to implement opaque predicates.
Domas~\cite{movfuscator} developed a compiler which generates a binary employing only the \texttt{mov} family
instructions, based on the fact that x86 \texttt{mov} is Turing complete.
There are other binary obfuscation methods which heavily relies on compression, encryption,
and virtualization~\cite{proceeding_Justin_RAID08,procecding_OmniUnpack_ACSAC07,procecding_PolyUnpack_ACSAC06}.
Among these obfuscation techniques, 
binary packers using compression and encryption can be vulnerable to dynamic analysis because
the original code has to be restored at some point of program execution. 
As for \vbo, most current approaches are implemented in the decode-dispatch scheme~\cite{smith05}. 
Recent effort~\cite{Rolles:2009:UVO:1855876.1855877,Sharif2009} has identified the
characteristics of the decode-dispatch pattern in the virtualization-obfuscated binaries so that
they can be effectively reverse engineered.

As for deobfuscation, most recent work focuses on attacking \vbo.
Sharif et al.~\cite{Sharif2009} has developed an outside-in approach which 
first reverse engineers the virtual machine and then decodes the bytecode to recover
the protected program.
%This approach heavily relies on some assumptions about the structure and working process of the virtual machine.
%If the virtualizer does meet these assumptions, the deobfuscator is likely to fail.
Another deobfuscation method presented by Coogan et al.~\cite{Coogan2011}
chooses the inside-out method
which utilizes equational reasoning to simplify the execution traces of protected programs. 
In this way, the deobfuscator extracts instructions which are truly relevant to program logic.
A very recent method proposed by Yadegari et al.~\cite{yadegari2014generic} improved the
inside-out approach with more generic control flow simplification 
algorithms that can deobfuscate programs protected by nested virtualization.
Without access to these tools, we cannot directly test \babel's resilience to them.
However, since \babel\ completely reforms C programs' data layout and reconstructs
the control flows with a much different programming paradigm,
we are very confident with \babel's security strength against these approaches.

Binary diffing is another widely used reverse engineering technique that takes
program obfuscation into account. Binary differs identify the syntactical or semantic
similarity between two different binaries, and can be used to detect programming plagiarism
and launch similarity-based attacks~\cite{brumley08}.
BinDiff~\cite{bindiff} and CoP~\cite{LUO}, the two differs we use for evaluating \babel's
resilience, are currently the state of the art.
Other examples of binary differs include DarunGrim2~\cite{dg2}, Bdiff~\cite{bdiff},
BinHunt~\cite{gao_binhunt:_2008}, and iBinHunt~\cite{iBinhunt}.
Although these tools can defeat certain types of program obfuscation,
none of them are designed to handle the complexity of \mo.

\section{Conclusion}\label{sec:conclusion}
In this paper, we present \mo, a new software obfuscation scheme based on translations from one programming language to another. By utilizing certain design and implementation
features of the target language, we are able to protect the original program against reverse engineering.
We implement \babel, a tool that translates part of a C program into Prolog 
and utilizes Prolog's unique language features to make the program obscure.
We evaluate \babel\ with respect to
potency, resilience, cost, and stealth on real-world C programs of different categories.
The experimental results show that \mo\ is an adequate and practical software protection
technique.

\section*{Acknowledgments}
We thank Herbert Bos and the anonymous reviewers
for their valuable feedback which has greatly helped us improve the paper.
This research was supported in part by the National Science Foundation (NSF)
grants CNS-1223710 and CCF-1320605, and the Office of
Naval Research (ONR) grant N00014-13-1-0175.

\bibliographystyle{IEEEtranS}
\bibliography{main,smartphone,wu}

\begin{appendix}
\section{Additional Potency Evaluation Data}
\label{sec:extra}
In this appendix, we present the program complexity at obfuscation levels of 10\%, 20\%, 30\%, 40\%, and 50\%, for both \babel\ (Table~\ref{tab:babel-potency-all}) and Code Virtualizer (Table~\ref{tab:cv-potency-all}). 
We would like to remind readers that for the \babel\ potency data, all values are obtained by IDA Pro~\cite{ida}.
Since \babel\ generates many indirect control flow (see \S~\ref{sec:oc-features} and \F~\ref{fig:example}),
it is hard to evaluate how accurate IDA Pro is. Nevertheless, in our case
the reported values can be interpreted as lower bounds of the corresponding metrics.
\begin{table*}
\centering
\caption{Program Complexity of \babel-Obfuscated Binaries at Different Obfuscation Levels}
\label{tab:babel-potency-all}
\begin{tabular}{l|c|P{4em}P{2.2em}|P{4em}P{2.2em}|P{4em}P{2.2em}|P{4em}P{2.2em}|P{4em}P{2.2em}}
\hline
\multirow{2}{*}{Program} & Obfuscation & \multicolumn{2}{c|}{\# of Call Graph Edges} & \multicolumn{2}{c|}{\# of CFG Edges} & \multicolumn{2}{c|}{\# of Basic Blocks} & \multicolumn{2}{c|}{Cyclomatic Number} & \multicolumn{2}{c}{Knot Count}\\
\cline{3-12}
& Level& Value & Ratio & Value & Ratio & Value & Ratio & Value & Ratio & Value & Ratio \\
\hline
\multirow{6}{*}{bzip2} & 0\%  & 353 & 1.0 & 5382 & 1.0 & 3528 & 1.0 & 1856 & 1.0 & 3120 & 1.0 \\
                          & 10\% & 5609 & 15.9 & 18539 & 3.4 & 15445 & 4.4 & 3096 & 1.7 & 12488 & 4.0 \\
                          & 20\% & 5719 & 16.2 & 18909 & 3.5 & 15788 & 4.5 & 3123 & 1.7 & 12166 & 3.9 \\
                          & 30\% & 5964 & 16.9 & 19771 & 3.7 & 17078 & 4.8 & 2695 & 1.5 & 12396 & 4.0 \\
                          & 40\% & 6386 & 18.1 & 19630 & 3.6 & 17907 & 5.1 & 1725 & 0.9 & 12027 & 3.9 \\
                          & 50\% & 6617 & 18.7 & 19829 & 3.7 & 18210 & 5.2 & 1621 & 0.9 & 12110 & 3.9 \\
\hline
\multirow{6}{*}{mcf} & No Obf.  & 78 & 1.0 & 854 & 1.0 & 583 & 1.0 & 273 & 1.0 & 153 & 1.0 \\
                          & 10\% & 5159 & 66.1 & 13352 & 15.6 & 11759 & 20.2 & 1595 & 5.8 & 8761 & 57.3 \\
                          & 20\% & 5302 & 68.0 & 13500 & 15.8 & 12079 & 20.7 & 1423 & 5.2 & 8761 & 57.3 \\
                          & 30\% & 5449 & 69.9 & 14233 & 16.7 & 13086 & 22.4 & 1149 & 4.2 & 8792 & 57.5 \\
                          & 40\% & 5519 & 70.8 & 13922 & 16.3 & 12926 & 22.2 & 998 & 3.7 & 8739 & 57.1 \\
                          & 50\% & 5697 & 73.0 & 14076 & 16.5 & 13464 & 23.1 & 614 & 2.2 & 8686 & 56.8 \\
\hline
\multirow{6}{*}{regexp} & No Obf.  & 72 & 1.0 & 855 & 1.0 & 591 & 1.0 & 266 & 1.0 & 1135 & 1.0 \\
                          & 10\% & 5053 & 70.2 & 13082 & 15.3 & 11447 & 19.4 & 1637 & 6.2 & 9675 & 8.5 \\
                          & 20\% & 5101 & 70.8 & 12964 & 15.2 & 11428 & 19.3 & 1538 & 5.8 & 9491 & 8.4 \\
                          & 30\% & 5276 & 73.3 & 13290 & 15.5 & 11802 & 20.0 & 1490 & 5.6 & 9530 & 8.4 \\
                          & 40\% & 5309 & 73.7 & 13064 & 15.3 & 11704 & 19.8 & 1362 & 5.1 & 9405 & 8.3 \\
                          & 50\% & 5375 & 74.7 & 13292 & 15.5 & 11940 & 20.2 & 1354 & 5.1 & 9393 & 8.3 \\
\hline
\multirow{6}{*}{svm} & No Obf.  & 511 & 1.0 & 5375 & 1.0 & 3545 & 1.0 & 1832 & 1.0 & 2972 & 1.0 \\
                          & 10\% & 5734 & 11.2 & 19156 & 3.6 & 15777 & 4.5 & 3381 & 1.8 & 11729 & 3.9 \\
                          & 20\% & 6343 & 12.4 & 19912 & 3.7 & 17368 & 4.9 & 2546 & 1.4 & 11658 & 3.9 \\
                          & 30\% & 6739 & 13.2 & 20752 & 3.9 & 18533 & 5.2 & 2221 & 1.2 & 11521 & 3.9 \\
                          & 40\% & 7052 & 13.8 & 20680 & 3.8 & 19049 & 5.4 & 1633 & 0.9 & 11547 & 3.9 \\
                          & 50\% & 7661 & 15.0 & 21119 & 3.9 & 20135 & 5.7 & 986 & 0.5 & 11552 & 3.9 \\
\hline
\multirow{6}{*}{oftpd} & No Obf.  & 455 & 1.0 & 2035 & 1.0 & 1667 & 1.0 & 370 & 1.0 & 1277 & 1.0 \\
                          & 10\% & 5541 & 12.2 & 14591 & 7.2 & 13011 & 7.8 & 1582 & 4.3 & 9856 & 7.7 \\
                          & 20\% & 5710 & 12.5 & 15110 & 7.4 & 13812 & 8.3 & 1300 & 3.5 & 9923 & 7.8 \\
                          & 30\% & 5810 & 12.8 & 15501 & 7.6 & 14422 & 8.7 & 1081 & 2.9 & 9911 & 7.8 \\
                          & 40\% & 5853 & 12.9 & 15875 & 7.8 & 15086 & 9.0 & 791 & 2.1 & 9858 & 7.7 \\
                          & 50\% & 6048 & 13.3 & 16493 & 8.1 & 16108 & 9.7 & 387 & 1.0 & 9954 & 7.8 \\
\hline
\multirow{6}{*}{mongoose} & No Obf.  & 1027 & 1.0 & 2788 & 1.0 & 2086 & 1.0 & 704 & 1.0 & 493 & 1.0 \\
                          & 10\% & 6288 & 6.1 & 15981 & 5.7 & 14262 & 6.8 & 1721 & 2.4 & 9495 & 19.3 \\
                          & 20\% & 6525 & 6.4 & 16464 & 5.9 & 15102 & 7.2 & 1364 & 1.9 & 9474 & 19.2 \\
                          & 30\% & 6762 & 6.6 & 17115 & 6.1 & 16079 & 7.7 & 1038 & 1.5 & 9491 & 19.3 \\
                          & 40\% & 6784 & 6.6 & 17597 & 6.3 & 16924 & 8.1 & 675 & 1.0 & 9447 & 19.2 \\
                          & 50\% & 7024 & 6.8 & 18470 & 6.6 & 18369 & 8.8 & 103 & 0.1 & 9450 & 19.2 \\
\hline
\end{tabular}
\end{table*}

\begin{table*}
\centering
\caption{Program Complexity of CV-Obfuscated Binaries at Different Obfuscation Levels}
\label{tab:cv-potency-all}
\begin{tabular}{l|c|P{4em}P{2.2em}|P{4em}P{2.2em}|P{4em}P{2.2em}|P{4em}P{2.2em}|P{4em}P{2.2em}}
\hline
\multirow{2}{*}{Program} & Obfuscation & \multicolumn{2}{c|}{\# of Call Graph Edges} & \multicolumn{2}{c|}{\# of CFG Edges} & \multicolumn{2}{c|}{\# of Basic Blocks} & \multicolumn{2}{c|}{Cyclomatic Number} & \multicolumn{2}{c}{Knot Count}\\
\cline{3-12}
& Level& Value & Ratio & Value & Ratio & Value & Ratio & Value & Ratio & Value & Ratio \\
\hline
\multirow{6}{*}{bzip2} & 0\%  & 353 & 1.0 & 5382 & 1.0 & 3528 & 1.0 & 1856 & 1.0 & 3120 & 1.0 \\
                          & 10\% & 424 & 1.2 & 4079 & 0.8 & 2962 & 0.8 & 1119 & 0.6 & 645 & 0.2 \\
                          & 20\% & 385 & 1.1 & 3988 & 0.7 & 2906 & 0.8 & 1084 & 0.6 & 630 & 0.2 \\
                          & 30\% & 261 & 0.7 & 3868 & 0.7 & 2826 & 0.8 & 1044 & 0.6 & 713 & 0.2 \\
                          & 40\% & 248 & 0.7 & 3675 & 0.7 & 2684 & 0.8 & 993 & 0.5 & 699 & 0.2 \\
                          & 50\% & 242 & 0.7 & 3652 & 0.7 & 2684 & 0.8 & 970 & 0.5 & 696 & 0.2 \\
\hline
\multirow{6}{*}{mcf} & No Obf.  & 78 & 1.0 & 854 & 1.0 & 583 & 1.0 & 273 & 1.0 & 153 & 1.0 \\
                          & 10\% & 36 & 0.5 & 531 & 0.6 & 377 & 0.6 & 156 & 0.6 & 71 & 0.5 \\
                          & 20\% & 36 & 0.5 & 520 & 0.6 & 370 & 0.6 & 152 & 0.6 & 71 & 0.5 \\
                          & 30\% & 34 & 0.4 & 461 & 0.5 & 329 & 0.6 & 134 & 0.5 & 68 & 0.4 \\
                          & 40\% & 24 & 0.3 & 372 & 0.4 & 270 & 0.5 & 104 & 0.4 & 54 & 0.4 \\
                          & 50\% & 33 & 0.4 & 308 & 0.4 & 223 & 0.4 & 87 & 0.3 & 46 & 0.3 \\
\hline
\multirow{6}{*}{regexp} & No Obf.  & 72 & 1.0 & 855 & 1.0 & 591 & 1.0 & 266 & 1.0 & 1135 & 1.0 \\
                          & 10\% & 69 & 1.0 & 589 & 0.7 & 410 & 0.7 & 181 & 0.7 & 619 & 0.5 \\
                          & 20\% & 67 & 0.9 & 578 & 0.7 & 411 & 0.7 & 169 & 0.6 & 618 & 0.5 \\
                          & 30\% & 66 & 0.9 & 525 & 0.6 & 377 & 0.6 & 150 & 0.6 & 603 & 0.5 \\
                          & 40\% & 58 & 0.8 & 330 & 0.4 & 243 & 0.4 & 89 & 0.3 & 117 & 0.1 \\
                          & 50\% & 57 & 0.8 & 326 & 0.4 & 253 & 0.4 & 75 & 0.3 & 117 & 0.1 \\
\hline
\multirow{6}{*}{svm} & No Obf.  & 511 & 1.0 & 5375 & 1.0 & 3545 & 1.0 & 1832 & 1.0 & 2972 & 1.0 \\
                          & 10\% & 403 & 0.8 & 3612 & 0.7 & 2576 & 0.7 & 1038 & 0.6 & 324 & 0.1 \\
                          & 20\% & 382 & 0.7 & 3431 & 0.6 & 2460 & 0.7 & 973 & 0.5 & 313 & 0.1 \\
                          & 30\% & 357 & 0.7 & 3267 & 0.6 & 2358 & 0.7 & 911 & 0.5 & 302 & 0.1 \\
                          & 40\% & 337 & 0.7 & 3129 & 0.6 & 2259 & 0.6 & 872 & 0.5 & 293 & 0.1 \\
                          & 50\% & 311 & 0.6 & 2971 & 0.6 & 2147 & 0.6 & 826 & 0.5 & 282 & 0.1 \\
\hline
\multirow{6}{*}{oftpd} & No Obf.  & 455 & 1.0 & 2035 & 1.0 & 1667 & 1.0 & 370 & 1.0 & 1277 & 1.0 \\
                          & 10\% & 444 & 1.0 & 1923 & 0.9 & 1582 & 0.9 & 343 & 0.9 & 1097 & 0.9 \\
                          & 20\% & 411 & 0.9 & 1786 & 0.9 & 1454 & 0.9 & 334 & 0.9 & 1065 & 0.8 \\
                          & 30\% & 390 & 0.9 & 1727 & 0.8 & 1435 & 0.9 & 294 & 0.8 & 542 & 0.4 \\
                          & 40\% & 333 & 0.7 & 1500 & 0.7 & 1237 & 0.7 & 265 & 0.7 & 972 & 0.8 \\
                          & 50\% & 307 & 0.7 & 1384 & 0.7 & 1158 & 0.7 & 228 & 0.6 & 944 & 0.7 \\
\hline
\multirow{6}{*}{mongoose} & No Obf.  & 1027 & 1.0 & 2788 & 1.0 & 2086 & 1.0 & 704 & 1.0 & 493 & 1.0 \\
                          & 10\% & 717 & 0.7 & 2489 & 0.9 & 1934 & 0.9 & 557 & 0.8 & 526 & 1.1 \\
                          & 20\% & 644 & 0.6 & 2239 & 0.8 & 1786 & 0.9 & 455 & 0.6 & 462 & 0.9 \\
                          & 30\% & 585 & 0.6 & 2063 & 0.7 & 1638 & 0.8 & 427 & 0.6 & 442 & 0.9 \\
                          & 40\% & 532 & 0.5 & 1954 & 0.7 & 1555 & 0.7 & 401 & 0.6 & 430 & 0.9 \\
                          & 50\% & 467 & 0.5 & 1787 & 0.6 & 1424 & 0.7 & 365 & 0.5 & 400 & 0.8 \\
\hline
\end{tabular}
\end{table*}

\end{appendix}
\end{document}